\documentclass[12pt,a4paper]{article}
\usepackage{amsmath,amssymb} %,tabularx
\usepackage{graphicx}
\usepackage{physics}
\usepackage{hyperref}% add hypertext capabilities
\usepackage{microtype}
\usepackage[title]{appendix}

\allowdisplaybreaks
\newcommand{\bea}{\begin{eqnarray}}
\newcommand{\eea}{\end{eqnarray}}
\newcommand{\sgn}{\text{sgn}}
\newcommand{\nn}{\nonumber} 
\DeclareMathOperator{\diag}{diag}
\newcommand{\bq}{\mathbf{q}} % for vector 
\newcommand{\ba}{\begin{align}}
\newcommand{\ea}{\end{align}}
\def\thetap{\theta^\prime_{e\mu}}

\begin{document}
\title{
\begin{flushright}
\ \\*[-120pt] 
\begin{minipage}{0.2\linewidth}
\normalsize
%arXiv:YYMM.NNNN \\
HUPD-2213 \\*[0pt]
\end{minipage}
\end{flushright}
{\Large 
Determination of Majorana type-phases \\ from
 the time evolution of lepton numbers 
\\*[5pt]} }
\author{\normalsize
\centerline{
Nicholas J. Benoit$^{1}$\footnote{E-mail address: njbenoit@hiroshima-u.ac.jp}, \,
Yuta Kawamura $^{3}$\footnote{E-mail address: kawamura1994phy@gmail.com}, \,
Saki Hamada $^{4}$\footnote{E-mail address: sakikwn@gmail.com}, \,
} \\ \normalsize
\centerline{
Takuya Morozumi $^{1,2}$\footnote{E-mail address: morozumi@hiroshima-u.ac.jp
}, \,
Yusuke Shimizu $^{5}$\footnote{E-mail address: shimizu.yusuke@kaishi-pu.ac.jp
}, \,
Kei Yamamoto $^{6}$\footnote{E-mail address: keiy@iwate-u.ac.jp}}
%%%%%%%%%%%%%%%%%%%%%%
\\*[0pt]
\centerline{
\begin{minipage}{\linewidth}
\begin{center} 
\vspace{0.3cm}
{\it \small
$^1$Physics Program Graduate School of Advanced Science and Engineering and $^2$Core of Research for the Energetic Universe,  Hiroshima University, 
Higashi-Hiroshima 739-8526, Japan} \\*[5pt]
{\it \small $^3$
Kitakami city, Iwate, Japan} \\*[5pt]
{\it \small $^4$
Higashi-Hiroshima, Japan} \\*[5pt]
{\it \small $^5$
Department~of~Information,~Kaishi~Professional~University, 
Niigata~950-0916,~Japan} \\*[5pt]
{\it \small $^6$
Faculty of Science and Engineering, 
Graduate school of Science and Engineering, 
Iwate University, 
 Morioka 020-0066, Japan} 
\vspace{-0.8cm}
\end{center}
\end{minipage}}
\\*[0pt]}
\date{}
\maketitle
\begin{abstract}
%\\*[50pt]}
%\date{
%\centerline{\small \bf Abstract}
%\begin{minipage}{0.9\linewidth}
%\medskip
%\medskip
%\small
We have investigated an approach for determining the Majorana type-phases using the time evolution of lepton family numbers.
We show how the second-order time derivative of the expectation values for the lepton family numbers depends on the sum of the Majorana type-phases.
Furthermore, others have connected the Majorana type-phases to the orientation of unitary triangles for the Pontecorvo-Maki-Nakagawa-Sakata (PMNS) matrix, and the usual Majorana phases $\alpha_{21}$ and $\alpha_{31}$.
Theoretically, this allows for the extraction of the orientation of the unitary triangles and the Majorana phases from lepton family numbers.
We study three example situations.
First, how to extract the Majorana type-phases and the lightest neutrino mass for three massive neutrinos, and when a neutrino is massless, i.e., $m_{1,3}=0$.
Second, the determination of the Majorana phase and the lightest neutrino mass for a two generation toy model.
Third, simplified realizations of the type I seesaw model with two gauge singlet neutrinos and two families of lepton doublets.
We calculate how the Majorana phases and Majorana type-phases are related to CP violation for leptogenesis at high energies.
At first, the effective Majorana mass matrix is parametrized with real and positive diagonal elements at low energies.
In this basis, the phase of the off-diagonal elements are related to the CP violating phases in the PMNS matrix.
We explicitly show the relation between the single Majorana phase and the phase of the effective Majorana mass matrix for the toy model with two generations of active neutrinos.
Then for the model with two gauge singlet neutrinos and two families of lepton doublets, we study the effective Majorana mass matrix generated by the seesaw model.
In that model, we can show how the Majorana phase at low energy is related to the two CP violating phases of the seesaw matrix.
That relation between the phases depends on the lepton number asymmetries of the heavy Majorana neutrinos decays for the toy models.
%\end{minipage}
%}
%\begin{titlepage}
%\maketitle
\end{abstract}
\thispagestyle{empty}
%\end{titlepage}
%%%%%%%%%%%%%%%%%%%%%%%%%%%
%%%%%%%%%%%%%%%%%%%%%%%%%%%%%%%%%%%%%%%%%%%%%%%%%%%%%%%%%%%% 
%%%%%%%%%%%%%%%%%%%%%%%%   Introduction   %%%%%%%%%%%%%%%%%%%%%%%%%%%
%%%%%%%%%%%%%%%%%%%%%%%%%%%%%%%%%%%%%%%%%%%%%%%%%%%%%%%%%%%%
\section{Introduction}
Despite phenomenal experimental programs in neutrino physics, open questions remain.
We know the three active neutrinos of the Standard Model come in three flavors (families) electron neutrino, muon neutrino, and the tauon neutrino~\cite{ALEPH:2005ab}.
Those flavors are assigned based on the neutrinos weak charged current lepton partner.
We also know that neutrinos can oscillate between flavors over macroscopic distances~\cite{Super-Kamiokande:1998kpq,SNO:2002tuh}.
To describe those oscillations, the three neutrino flavors are treated as linear combinations of massive eigenstates.
Then to explain experimental data, at least two of the eigenstates must have small, but non-zero, masses~\cite{Pontecorvo:1967fh,Gribov:1968kq}.
This contrasts with the Standard Model, in which the three neutrinos are massless particles.
Because neutrino flavor oscillation experiments can only constrain the mass squared differences of the neutrinos $\Delta m^2_{ij}=m^2_i-m^2_j$, details of neutrino masses remain unknown.
For example, neutrinos are neutrally charged particles with non-zero mass, which means they could be their own antiparticles.
Under that setup, neutrinos would be called Majorana particles and have a Majorana mass~\cite{Majorana:1937vz,Racah:1937qq,Furry:1938zz}.
A Dirac particle type and mass are also possible for neutrinos, this would be similar to the usual Standard Model leptons.
Neither Majorana nor Dirac particle type for neutrinos has been experimentally established, despite years of searching for Majorana neutrinos.
Perhaps the most famous experiments are searches for neutrinoless double-beta decay~\cite{Furry:1939qr,Schechter:1981bd,Nieves:1984sn,Takasugi:1984xr,Umehara:2008ru,GERDA:2020xhi,Majorana:2019nbd,Barabash:2010bd,NEMO-3:2015jgm,NEMO-3:2016mvr,NEMO-3:2016qxo,NEMO-3:2016zfx,KamLAND-Zen:2022tow,EXO-200:2019rkq, Dolinski:2019nrj}, but there are also neutrinoless quadruple-beta decay~\cite{NEMO-3:2017gmi}, neutrinoless double-electron capture~\cite{Winter:1955zz,Bernabeu:1983yb,Sujkowski:2003mb,CUORE:2022dwv,XMASS:2018txy,Blaum:2020ogl}, proposed processes mediated by the exchange of a pair of virtual neutrinos akin to a Casimir force~\cite{Grifols:1996fk,Segarra:2020rah,Costantino:2020bei}, coherent scattering of neutrinos on nuclei with bremsstrahlung~\cite{Millar:2018hkv}, and proposed searches of quantum statics for final state decay processes involving neutrinos~\cite{Nieves:1985ir,Chhabra:1992be,Kayser:1981nw,Kim:2021dyj}.

We will consider the situation that neutrinos are Majorana particles.
In sec.~\ref{sec:MajoranaTypePhases}, we summarize how the three neutrino flavors are constructed as linear combinations of the massive eigenstates via the PMNS mixing matrix.
We also introduce how the mixing matrix can be formed with Majorana type-phases and unitary triangles.
Then, in sec.~\ref{sec:timeevoMajoranatypephases}, we connect our previous work on the time evolution of lepton family numbers to the Majorana type-phases.
After assuming the lepton family numbers can be measured in the future, we give an example of how our previous work can be used to determine the Majorana type-phases and the lightest neutrino mass in sec.~\ref{sec:determineMajoranatypephasesandlightestmass}.
In sec.~\ref{sec:lightestismassless}, we consider the lightest neutrino to be massless and illustrate how this enhances the predictive power of our work.
Sections~\ref{sec:Meff} and \ref{sec:CPVseesaw} are devoted to relations between the effective Majorana mass matrix, Majorana phases for a toy model of two generations of active neutrinos, and CP violation of leptogenesis~\cite{Fukugita:1986hr} in the type I seesaw model~\cite{Minkowski:1977sc,Gell-Mann:1979vob,Mohapatra:1979ia,Yanagida:1980xy} with two heavy Majorana neutrinos.
Finally, we provide some concluding thoughts.

%%%%%%%%%%%%%%%%%%%%%%%%%%%%%%%%%%%%%%%%%%%%%%%%%%%%%%%%%%%%
%%%%%%%%%%%%%%%%%%%%%%%%   Body 1   %%%%%%%%%%%%%%%%%%%%%%%%%%%
%%%%%%%%%%%%%%%%%%%%%%%%%%%%%%%%%%%%%%%%%%%%%%%%%%%%%%%%%%%%
\section{Majorana type-phases and unitary triangles}\label{sec:MajoranaTypePhases}
Neutrino masses are a pillar for physics beyond the Standard Model that started with the discovery of neutrino flavor oscillations \cite{Abe:2010hy,Aharmim:2009gd}.
The results of neutrino oscillation experiments are mathematically described by a mixing of basis states via the Pontecorvo-Maki-Nakagawa-Sakata (PMNS) matrix \cite{Pontecorvo:1957qd,Maki:1962mu,Giunti:2007ry}
\begin{equation}
    \nu_{L\alpha}=U_{\alpha i}\nu_{Li}=
    \begin{pmatrix}
        U_{e1}&U_{e2}&U_{e3} \\
        U_{\mu 1}&U_{\mu 2}&U_{\mu 3} \\
        U_{\tau 1}&U_{\tau 2}&U_{\tau 3}
    \end{pmatrix}
    \begin{pmatrix}
        \nu_{L1}\\
        \nu_{L2}\\
        \nu_{L3}
    \end{pmatrix},
\end{equation}
where $\alpha = e,\mu,\tau$ and $i = 1,2,3$.
The PMNS matrix $U_{\alpha i}$ can be formed for neutrinos with a Majorana mass though diagonalization of the Majorana mass matrix,
\begin{equation}
    (U)_{\alpha i}(m_\nu)_{\alpha\beta}(U)_{\beta j} = (m_\nu)_i\delta_{ij} ,
\end{equation}
using an Autonne -Takagi Factorization \cite{Autonne:1915},\cite{Takagi:1925},\cite{Hahn:2006hr}.
Then the leptonic weak charged current interaction is written as,
\begin{equation}
    \mathcal{L}^{(CC)}_I = \frac{g}{\sqrt{2}}\overline{l_{L\alpha}}\gamma_{\mu}U_{\alpha i}\nu_{Li}W^{\mu-}+\text{h.c.}\,.
\end{equation}
The charged leptons are free to be re-phased, because their mass term is invariant under the phase transformations of $l_\alpha \rightarrow l^{'}_\alpha=\exp(i\phi_\alpha)l_\alpha$.
But, the neutrino Majorana mass term is not invariant under phase transformations.
This means the standard parametrization of the PMNS matrix \cite{Zyla:2020zbs} must be extended to include two independent CP violating phases.
Thus, the unitary $3\times3$ PMNS matrix depends on three mixing angles and three CP violating phases.
\begin{equation}
   U= U^D\diag\left(1,\,e^{i \frac{\alpha_{21}}{2}},\,e^{i\frac{\alpha_{31}}{2}} \right),
    \label{eq:PMNSmatrix}
\end{equation}
where the Dirac portion $U^D$ is of the form,
\begin{equation}
    U^D =
    \begin{pmatrix}
        c_{12}c_{13} & s_{12}c_{13} & s_{13}e^{-i\delta}\\
        -s_{12}c_{23}-c_{12}s_{23}s_{13}e^{i\delta} & c_{12}c_{23}-s_{12}s_{23}s_{13}e^{i\delta} & s_{23}c_{13} \\
        s_{12}s_{23}-c_{12}c_{23}s_{13}e^{i\delta} & -c_{12}s_{23}-s_{12}c_{23}s_{13}e^{i\delta} & c_{23}c_{13}
    \end{pmatrix}.
    \label{eq:DiracPMNSmatrix}
\end{equation}
We use the usual notation of $c_{ij}=\cos\theta_{ij}$ and $s_{ij}=\sin\theta_{ij}$ from PDG~\cite{Zyla:2020zbs}.
The CP violating phases in the diagonal matrix, $\alpha_{21}$ and $\alpha_{31}$, are the Majorana phases~\cite{Bilenky:1980cx,Schechter:1980gr,Doi:1980yb}, whereas the CP violating phase of $U^D$ is the Dirac Kobayashi-Maskawa phase $\delta$ \cite{Kobayashi:1973fv}.
The CP violating phases have physical meaning only if all the mixing angles are not $0$ or $\pi/2$.
This can be seen by constructing the invariants~\cite{Jarlskog:1985ht,Branco:1986gr,Nieves:1987pp,Feruglio:2012cw} of the PMNS matrix.

We consider an interesting aspect of Majorana neutrinos and the rephasing invariant bilinears~\cite{Nieves:1987pp} that Branco and Rebelo named ``Majorana type-phases''~\cite{Branco:2008ai}.
The Majorana type-phases are defined to be the argument of the bilinears $U_{\alpha k}U^\ast_{\alpha j}$ with no summation over the repeated index $\alpha$.
Then, Branco and Rebelo choose six Majorana type-phases and prove the six phases can be used to exactly reproduce the $3\times3$ PMNS matrix Eq.(\ref{eq:PMNSmatrix}).
We reproduce their~\cite{Branco:2008ai} six Majorana type-phases for self-containment of our work,
\begin{align} \label{eq:majoranaTypeBeta}
    \beta_1 = \arg{U_{e1}U^\ast_{e2}}\,, && \beta_2 = \arg{U_{\mu 1}U^\ast_{\mu 2}}\,, && \beta_3 = \arg{U_{\tau 1}U^\ast_{\tau 2}}\,,
  \\ \label{eq:majoranaTypeGamma}
    \gamma_1 = \arg{U_{e1}U^\ast_{e3}}\,, && \gamma_2 = \arg{U_{\mu 1}U^\ast_{\mu 3}}\,, && \gamma_3 = \arg{U_{\tau 1}U^\ast_{\tau 3}}\,.
\end{align}

\begin{figure}[htp]
    \centering
\includegraphics[]{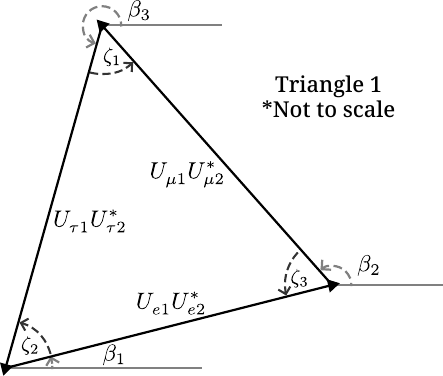}
    \caption{\label{fig:majoranatri1}
    The first Majorana triangle depends on the Majorana type-phases $\beta_k$.
    The sides of the triangle are constructed from the first two rows of the PMNS matrix.
    The orientation is physically meaningful and can only be determined by knowledge of the Majorana type-phases.}
\end{figure}

To recreate the $3\times 3$ PMNS matrix of Eq.(\ref{eq:PMNSmatrix}), from the Majorana type-phases Eq.(\ref{eq:majoranaTypeBeta}) and Eq.(\ref{eq:majoranaTypeGamma}), we can consider the construction of unitary triangles.
Two types of triangles can be created from the PMNS matrix, Dirac and Majorana triangles.
We are interested in Majorana triangles because of their dependence on the Majorana type-phases.
The three Majorana triangles are derived by multiplying the columns of the PMNS matrix,
\begin{align} \label{eq:majoranatriangle1}
    U^{}_{e1}U^\ast_{e2}+U^{}_{\mu 1}U^\ast_{\mu 2}+U^{}_{\tau 1}U^\ast_{\tau 2} &=0, && \text{triangle 1,}
    \\ \label{eq:majoranatriangle2}
    U^{}_{e1}U^\ast_{e3}+U^{}_{\mu 1}U^\ast_{\mu 3}+U^{}_{\tau 1}U^\ast_{\tau 3} &=0, && \text{triangle 2,}
    \\ \label{eq:majoranatriangle3}
    U^{}_{e2}U^\ast_{e3}+U^{}_{\mu 2}U^\ast_{\mu 3}+U^{}_{\tau 2}U^\ast_{\tau 3} &=0, && \text{triangle 3.}
\end{align}
The equation for triangle 1, Eq.(\ref{eq:majoranatriangle1}), is connected with the $\beta_k$ Majorana type-phases of Eq.(\ref{eq:majoranaTypeBeta}).
We illustrate that connection in Fig.~\ref{fig:majoranatri1} by plotting the triangle vector of Eq.(\ref{eq:majoranatriangle1}) denoting the angles $\beta_k$.
Additionally, we can see all sides of the triangle are constructed from the rephasing invariants $U_{\alpha k}U^\ast_{\alpha j}$.
This means the Majorana triangles are not free to rotate in the complex plain and their orientation is physically meaningful~\cite{Branco:2008ai,Aguilar-Saavedra:2000jom}.
The internal angles of the triangle $\zeta_k$ can be calculated as,
\begin{align}
    \zeta_1 = \pi - \left( \beta_3-\beta_2 \right), &&
    \zeta_2 = \left( \beta_3-\beta_1 \right)-\pi, &&
    \zeta_3 = \pi - \left( \beta_2-\beta_1 \right). \label{eq:internal}
\end{align}
Triangle 2 is connected to the $\gamma_k$ of Eq.(\ref{eq:majoranaTypeGamma}) and behaves similarly to triangle 1 where the orientation is physically meaningful.
The internal angles of triangle 2 are calculated by replacing $\beta_k$ with $\gamma_k$ in Eq.(\ref{eq:internal}).
Lastly, the connection to triangle 3 comes from subtraction between the $\beta_k$'s and $\gamma_k$'s and its behavior depends on the other two triangles.

Next, Branco and Rebelo related the mixing angles of the parametrized PMNS matrix in Eq.(\ref{eq:DiracPMNSmatrix}) to the Majorana type-phases using the law of sines on the Majorana triangles.
For example, to find $\theta_{12}$ they take,
\begin{equation} \label{eq:theta12}
    \begin{split}
        \tan^2\theta_{12}&{}=\frac{\abs{U_{e2}U^\ast_{e3}}^2}{\abs{U_{e1}U^\ast_{e3}}^2}
        \\
        &{}=\frac{\abs{\sin(\gamma_1-\gamma_3)}\abs{\sin(\gamma_2-\gamma_1)}\abs{\sin(\gamma_3-\gamma_2-(\beta_3-\beta_2))}}{\abs{\sin(\gamma_1-\gamma_3-(\beta_1-\beta_3))}\abs{\sin(\gamma_2-\gamma_1-(\beta_2-\beta_1))}\abs{\sin(\gamma_3-\gamma_2)}}.
    \end{split}
\end{equation}
Then they prove how the Dirac phase $\delta$ relates to the Majorana type-phases via the common area of the triangles,
\begin{equation} \label{eq:deltacp}
    \begin{split}
        A&{}=\frac{1}{16}\abs{\sin2\theta_{12}\sin2\theta_{13}\sin2\theta_{23}\cos\theta_{13}\sin\delta} \\
        &{}=\frac{1}{2}\abs{\cos\theta_{12}\cos\theta_{13}\sin\theta_{13}}^2\frac{\abs{\sin(\gamma_1-\gamma_2)}\abs{\sin(\gamma_1-\gamma_3)}}{\abs{\sin(\gamma_2-\gamma_3)}}.
    \end{split}
\end{equation}

\begin{figure}[htp]
    \begin{center}
    \includegraphics[width=\textwidth]{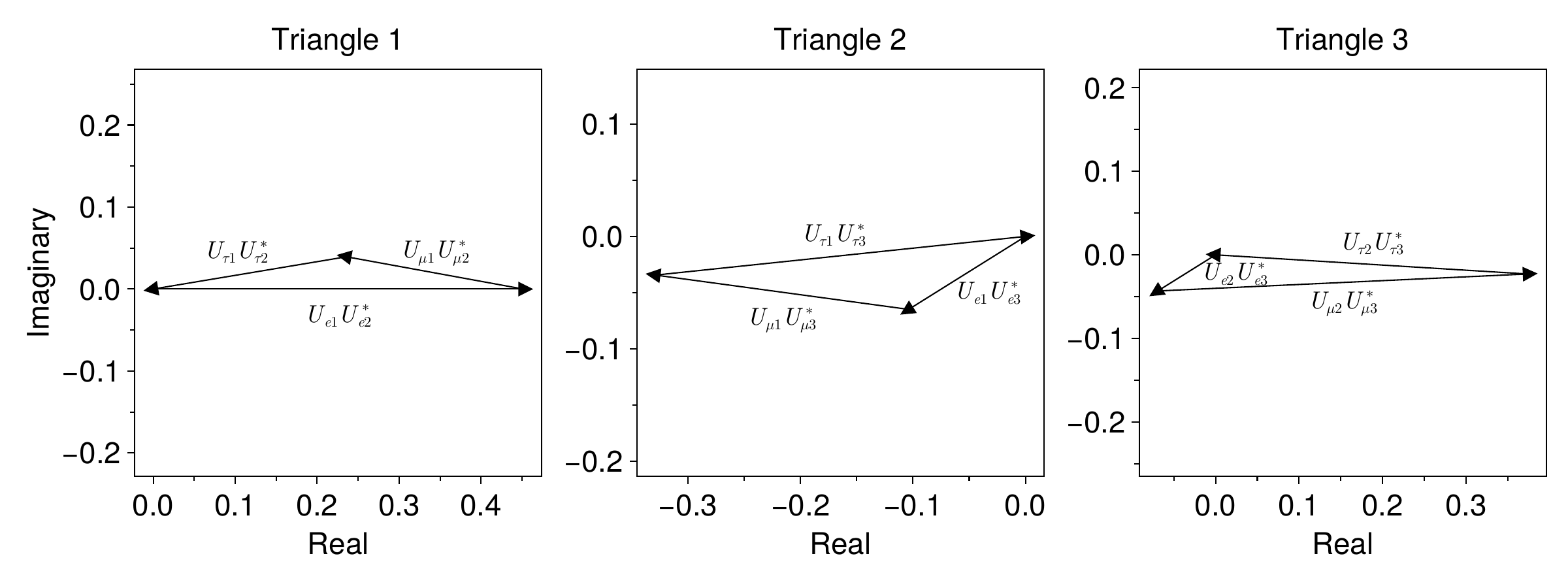}
    \caption{\label{fig:tripletriangles} We build the three Majorana triangle using the best fit values of NuFITv6.0\cite{Esteban:2024eli} for normal hierarchy.
    The orientation of the triangles depends on the Majorana type-phases; $\beta_k$ for triangle 1 that is related to the Majorana phase $\alpha_{21}$, $\gamma_k$ for triangle 2, that is related to the Majorana phase $\alpha_{31}$, and for triangle 3
the subtraction between the Majorana type-phases, $\gamma_k-\beta_k$.
    For this instance, we set the Majorana phase $\alpha_{21}=\alpha_{31}=0$, because it is experimentally not determined.
    NuFITv6.0 assumes the PMNS matrix is unitary, so all triangles are completely closed.}
    \end{center}
\end{figure}

We use the relations like Eq.(\ref{eq:theta12}) and Eq.(\ref{eq:deltacp}) to update their analysis~\cite{Branco:2008ai} of the Majorana triangles based on the recent neutrino oscillation global fit.
Specifically, we use the best fit values of NuFITv6.0 calculated by the Nu-Fit collaboration~\cite{Esteban:2024eli}.
Compared to the analysis of Ref.~\cite{Branco:2008ai}, we include additional data from neutrino oscillation experiments.
For simplicity, we consider the Majorana phases to be absent $\alpha_{21}=\alpha_{31}=0$, and later we will include them.
The vectors for triangle 1 in Fig.~\ref{fig:tripletriangles} are,
\begin{equation}
  \begin{split}
    U^{}_{e 1}U^\ast_{e 2} &= 0.4514 - 0.0i,
    \\
    U^{}_{\mu 1}U^\ast_{\mu 2} &= -0.2156 + 0.03936i,
    \\
    U^{}_{\tau 1}U^\ast_{\tau 2} &= -0.2357 - 0.03936i,
  \end{split}
\end{equation}
where we only write the first four significant figures with no rounding.
We treat the vectors for triangles 2 and 3 of Fig.~\ref{fig:tripletriangles} similarly as triangle 1,
\begin{gather}
  \left.\begin{aligned}
    U^{}_{e1}U^\ast_{e3} &= -0.1038 - 0.06487i, \\
    U^{}_{\mu 1}U^\ast_{\mu 3} &= -0.2251 + 0.03049i, \\
    U^{}_{\tau 1}U^\ast_{\tau 3} &= 0.3289 + 0.03438i;
    \end{aligned}
  \right\}\text{Triangle 2}
\\
  \left.\begin{aligned}
    U^{}_{e2}U^\ast_{e3} &= -0.06926 - 0.04328i, \\
    U^{}_{\mu 2}U^\ast_{\mu 3} &= 0.4431 + 0.02034i, \\
    U^{}_{\tau 2}U^\ast_{\tau 3} &= -0.3738 + 0.02293i.
    \end{aligned}
  \right\}\text{Triangle 3}
\end{gather}
For all three panels in Fig~\ref{fig:tripletriangles} the vectors completely close to form a triangle.
This is because the data we are using from NuFITv6.0 assumes the PMNS matrix Eq.(\ref{eq:DiracPMNSmatrix}) is unitary.
If the unitary assumption was relaxed, the vectors could form an open triangle instead.
Furthermore, all three triangles are scalene, which is different from the isosceles triangles of Branco and Rebelo because we do not consider perturbations around tri-bimaximal mixing.

\begin{figure}[htp]
    \begin{center}
    \includegraphics[width=0.5\textwidth]{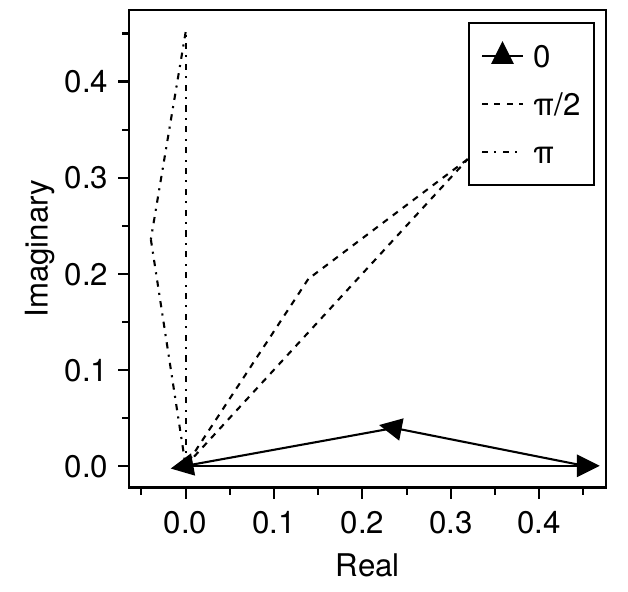}
    \caption{\label{fig:triangle1rotates} We rotate the first Majorana triangle from the best fit values of NuFITv6.0\cite{Esteban:2024eli} for normal hierarchy.
    The orientation of the triangle depends on the Majorana type-phases $\beta_k$, thus it is rotated by varying between $0<\alpha_{21}<\pi$.}
    \end{center}
\end{figure}

Lastly, current and near future neutrino oscillation experiments can not determine the orientation of the Majorana triangles.
This is because the oscillation experiments measure the Dirac phase $\delta$, but can not measure the Majorana phases $\alpha_{21}$ and $\alpha_{31}$ that determine the triangle orientation.
For example, we change the values of $\Delta \alpha_{21}$ to rotate triangle 1 in Fig.~\ref{fig:triangle1rotates}.
That rotation can not be fixed by current and near future neutrino oscillation experiments.
Branco and Rebelo discussed this problem~\cite{Branco:2008ai}, stating that measurements of the Dirac phase can only provide insight to the differences of the Majorana type-phases, i.e.$\gamma_1-\gamma_3$; so the triangle orientation can not be determined.
Whereas, the sum of Majorana type-phases $\beta_i+\beta_j$ and $\gamma_i+ \gamma_j$ are connected to the Majorana phases $\alpha_{21}$ and $\alpha_{31}$, respectively.
They then go on to discuss the possibility of determining the triangle orientation with neutrinoless double-beta decay and flavor sensitive leptogenesis~\cite{Fujihara:2005pv,Endoh:2003mz}.
In this paper, we will describe an additional method that can theoretically determine the triangle orientation.
This method is primarily based on our previous work of lepton number oscillations~\cite{Adam:2021qiq,Benoit:2022dsc,SalimAdam:2021suq}. 

%%%%%%%%%%%%%%%%%%%%%%%%%%%%%%%%%
\section{Time evolution of lepton number and Majorana type-phases}\label{sec:timeevoMajoranatypephases}
To determine the triangle orientation, we study the time evolution of lepton number focusing on the dependence of the Majorana type-phases.
In the works~\cite{Adam:2021qiq,Benoit:2022dsc,SalimAdam:2021suq} the lepton family numbers $L_\alpha^M$ for the Majorana neutrinos were defined.
Then, the time evolution of their expectation values were obtained.
We rewrite the result for the Majorana lepton numbers,
\begin{multline} \label{eq:lepton_M}
  \bra*{\sigma}L_\alpha^M(t)\ket*{\sigma} \\
  = \sum_{i,j}^3 \left[ \Re(U^\ast_{\alpha i} U^{}_{\sigma i} U^{}_{\alpha j} U^\ast_{\sigma j})\left( \cos\{E_i(\bq)t\} \cos\{E_j(\bq)t\} + \frac{\bq^2}{E_i(\bq) E_j(\bq)} \sin\{E_i(\bq)t\} \sin\{E_j(\bq)t\} \right)\right. \\
  - \Im(U^\ast_{\alpha i} U^{}_{\sigma i} U^{}_{\alpha j} U^\ast_{\sigma j})\left( \frac{\abs{\bq}}{E_i(\bq)} \sin\{E_i(\bq)t\} \cos\{E_j(\bq)t\} -\frac{\abs{\bq}}{E_j(\bq)} \cos\{E_i(\bq)t\} \sin\{E_j(\bq)t\} \right) \\
  \left.- \Re(U^\ast_{\alpha i} U^\ast_{\sigma i} U^{}_{\alpha j} U^{}_{\sigma j}) \frac{m_i}{E_i(\bq)}\frac{m_j}{E_j(\bq)} \sin\{E_i(\bq)t\} \sin\{E_j(\bq)t\} \right].
\end{multline}
In our notation the initial state $\sigma$ denotes a neutrino with a definite flavor i.e., $\sigma=e,\mu,\tau$ and momentum $\bq$.
The energy of the mass states we have defined as $E_{i,j}(\bq)=\sqrt{\bq^2+m_{i,j}^2}$.

To extract the Majorana type-phases we are interested in the third term of Eq.(\ref{eq:lepton_M}) that depends on the PMNS combination of $\Re(U^\ast_{\alpha i} U^\ast_{\sigma i} U^{}_{\alpha j} U^{}_{\sigma j})$.
To isolate the third term, consider that the first two terms can be completely determined from neutrino oscillation and neutrino mass experiments.
This means we can subtract those two terms from the Majorana expectation value to isolate the third term.
We define a new quantity $L^Q_\alpha(t)$ which denotes the sum of the first and the second term of Eq.(\ref{eq:lepton_M}),
\begin{multline} \label{eq:lepton_D}
  \bra*{\sigma}{L^Q}_\alpha(t)\ket*{\sigma} \\
   =  \sum_{i,j}^3 \left[ \Re(U^\ast_{\alpha i} U^{}_{\sigma i} U^{}_{\alpha j} U^\ast_{\sigma j})\left( \cos\{E_i(\bq)t\} \cos\{E_j(\bq)t\} + \frac{\bq^2}{E_i(\bq) E_j(\bq)} \sin\{E_i(\bq)t\} \sin\{E_j(\bq)t\} \right) \right.\\
   \left.- \Im(U^\ast_{\alpha i} U^{}_{\sigma i} U^{}_{\alpha j} U^\ast_{\sigma j})\left( \frac{\abs{\bq}}{E_i(\bq)} \sin\{E_i(\bq)t\} \cos\{E_j(\bq)t\} -\frac{\abs{\bq}}{E_j(\bq)} \cos\{E_i(\bq)t\} \sin\{E_j(\bq)t\} \right)\right],
\end{multline}
which has been determined from neutrino oscillation and neutrino mass experiments.
Assuming the expectation value of lepton family number in Eq.(\ref{eq:lepton_M}) can be experimentally measured, we denote the subtraction between it and the quantity of Eq.(\ref{eq:lepton_D}) as,
\begin{equation}
    \bra*{\sigma}L_\alpha^{M-Q}(t)\ket*{\sigma} \equiv \bra*{\sigma}L_\alpha^{M}(t)\ket*{\sigma}-\bra*{\sigma}L_\alpha^{Q}(t)\ket*{\sigma}.
\end{equation}
This value is the isolated third term of Eq.(\ref{eq:lepton_M}),
\begin{equation}
  \begin{split}\label{eq:majoranaDiracDiff}
    \bra*{\sigma}L_\alpha^{M-Q}(t)\ket*{\sigma} ={}&-\sum_{i,j}^3 \Re(U^\ast_{\alpha i} U^\ast_{\sigma i} U^{}_{\alpha j} U^{}_{\sigma j}) \frac{m_i}{E_i(\bq)}\frac{m_j}{E_j(\bq)} \sin\{E_i(\bq)t\} \sin\{E_j(\bq)t\},
\\
    ={}& -\sum_{i=1}^3 \abs{U_{\alpha i}U_{\sigma i}}^2 \left(\frac{m_i \sin\{E_i(\bq)t\}}{E_i(\bq)}\right)^2 
\\
    &-2  \sum_{i<j} \Re(U^\ast_{\alpha i} U^{}_{\alpha j} U^\ast_{\sigma i} U^{}_{\sigma j}) \frac{m_i}{E_i(\bq)}\frac{m_j}{E_j(\bq)} \sin\{E_i(\bq)t\} \sin\{E_j(\bq)t\}.
  \end{split}
\end{equation}
Let us investigate that equation by focusing on the dependence of the Majorana type-phases from Eq.(\ref{eq:majoranaTypeBeta}).
We take $\sigma=\alpha=e$ to obtain,
\begin{equation}
  \begin{split}\label{eq:m-d}
    \bra*{e}L_e^{M-Q}(t)\ket*{e}={}&-\sum_{i=1}^3 \abs{U_{e i}}^4 \left(\frac{m_i\sin\{E_i(\bq)t\}}{E_i(\bq)}\right)^2  
\\
    &-2 \Re\{ (U^\ast_{e 1}  U^{}_{e 2})^2\} \frac{m_1 \sin\{E_1(\bq)t\} }{E_1(\bq)}\frac{m_2  \sin\{E_2(\bq)t\} }{E_2(\bq)}
\\
    &-2 \Re \{ (U^\ast_{e 2} U^{}_{e 3})^2 \} \frac{m_2 \sin\{E_2(\bq)t\} }{E_2(\bq)}\frac{m_3 \sin\{E_3(\bq)t\} }{E_3(\bq)}
\\
    &-2 \Re \{(U^\ast_{e 1} U^{}_{e 3})^2 \} \frac{m_1 \sin\{E_1(\bq)t\} }{E_1(\bq)}\frac{m_3 \sin\{E_3(\bq)t\} }{E_3(\bq)}.
  \end{split}
\end{equation}
In the last three terms of Eq.(\ref{eq:m-d}), the PMNS combinations can be written with the Majorana type-phases $ 2\beta_1$ and $2\gamma_1$ based on Eq.(\ref{eq:majoranaTypeBeta}) and Eq.(\ref{eq:majoranaTypeGamma}).
The explicit dependence of Eq.(\ref{eq:m-d}) on the Majorana type-phases is then,
\begin{equation}
  \begin{split} \label{eq:Lee}
    \bra*{e}L_e^{M-Q}(t)\ket*{e} ={}&-\sum_{i=1}^3 \abs{U_{e i}}^4 \left(\frac{m_i\sin\{E_i(\bq)t\}}{E_i(\bq)} \right)^2
\\
    &-2 \abs{U_{e1}U_{e2}}^2 \cos(2 \beta_1) \frac{m_1 \sin\{E_1(\bq)t\} }{E_1(\bq)}\frac{m_2  \sin\{E_2(\bq)t\} }{E_2(\bq)}
\\
    &-2 \abs{U_{e2}U_{e3}}^2 \cos(2 (\gamma_1- \beta_1)) \frac{m_2 \sin\{E_2(\bq)t\} }{E_2(\bq)}\frac{m_3 \sin\{E_3(\bq)t\} }{E_3(\bq)}
\\
    &-2 \abs{U_{e1}U_{e3}}^2 \cos(2 \gamma_1) \frac{m_1 \sin\{E_1(\bq)t\} }{E_1(\bq)}\frac{m_3 \sin\{E_3(\bq)t\} }{E_3(\bq)}.
  \end{split}
\end{equation}
We can obtain similar results as Eq.(\ref{eq:Lee}) for the muon and tauon numbers of $\alpha=\mu,\tau$,
\bea
  \begin{split}\label{eq:Lemu}
    \bra*{e}L_\mu^{M-Q}(t)\ket*{e}
    =&{}-\sum_{i=1}^3 \abs{U_{\mu i}U_{e i}}^2 \left(\frac{m_i \sin\{E_i(\bq)t\} }{E_i(\bq)}\right)^2
\\
    &-2 \abs{U^\ast_{e 1} U^{}_{e 2} U^\ast_{\mu 1} U_{\mu 2}} \cos(\beta_1+\beta_2) \frac{m_1 \sin\{E_1(\bq)t\} }{E_1(\bq)}\frac{m_2 \sin\{E_2(\bq)t\} }{E_2(\bq)}
\\
    &-2 \abs{U^\ast_{e 2} U^{}_{e 3}  U^\ast_{\mu 2} U^{}_{\mu 3}}\cos(\gamma_1-\beta_1+\gamma_2-\beta_2) \frac{m_2 \sin\{E_2(\bq)t\} }{E_2(\bq)}\frac{m_3 \sin\{E_3(\bq)t\}}{E_3(\bq)}
\\
    & -2 \abs{U^\ast_{e 1} U^{}_{e 3}  U^\ast_{\mu 1} U^{}_{\mu 3}} \cos(\gamma_1+\gamma_2) \frac{m_1 \sin\{E_1(\bq)t\}}{E_1(\bq)}\frac{m_3 \sin\{E_3(\bq)t\}}{E_3(\bq)}.
  \end{split}
\eea
In contrast to Eq.(\ref{eq:Lee}), the muon number $L_\mu^{M-Q}(t)$ of Eq.(\ref{eq:Lemu}) depends on the Majorana type-phases $\beta_1+\beta_2$ and $\gamma_1+\gamma_2$.
Finally, the tauon number $L_\tau^{M-Q}(t)$ is written as,
\bea
  \begin{split}\label{eq:Letau}
    \bra*{e}L_\tau^{M-Q}(t)\ket*{e} 
    ={}& -\sum_{i=1}^3 \abs{U_{\tau i}U_{e i}}^2 \left(\frac{m_i \sin\{E_i(q)t\} }{E_i(q)}\right)^2 
\\
    &-2 \abs{U^\ast_{e 1} U^{}_{e2} U^\ast_{\tau 1} U^{}_{\tau 2}} \cos(\beta_1+\beta_3) \frac{m_1 \sin\{E_1(\bq)t\} }{E_1(\bq)}\frac{m_2 \sin\{E_2(\bq)t\} }{E_2(\bq)}
\\
    &-2 \abs{U^\ast_{e 2} U_{e 3} U^\ast_{\tau 2} U_{\tau3}} \cos(\gamma_1-\beta_1+\gamma_3-\beta_3) \frac{m_2 \sin\{E_2(\bq)t\}}{E_2(\bq)}\frac{m_3 \sin\{E_3(\bq)t\}}{E_3(\bq)}
 \\
    &-2 \abs{U^\ast_{e 1} U_{e 3} U^\ast_{\tau 1} U_{\tau 3}} \cos(\gamma_1+\gamma_3) \frac{m_1 \sin\{E_1(\bq)t\}}{E_1(\bq)}\frac{m_3 \sin\{E_3(\bq)t\} }{E_3(\bq)}.
  \end{split}
\eea
The last three terms are written with the Majorana type-phases $\beta_1+\beta_3$ and $\gamma_1+\gamma_3$.
All the lepton family numbers Eqs.(\ref{eq:Lee}-\ref{eq:Letau}) depend on the sum of Majorana type-phases $\beta_1+\beta_j$ and $\gamma_1+\gamma_j$ where $j=1,2,3$.  
In contrast to the difference between Majorana type-phases in an unitary triangle, such as $\beta_i-\beta_j$ and $\gamma_i-\gamma_j$, the sum of the Majorana type-phases depends on the Majorana phases $\alpha_{21}$ and $\alpha_{31}$ of Eq.(\ref{eq:PMNSmatrix}).

As we have seen in Eqs.(\ref{eq:Lee}-\ref{eq:Letau}), one can identify the part that depends on the sum of the Majorana type-phases, i.e., $\beta_i+\beta_j$ and $\gamma_i+\gamma_j$.
From that part, one may determine the summed value for the Majorana type-phases by fitting a curve to Eq.(\ref{eq:majoranaDiracDiff}) varying the unknown parameters.
A potential drawback of this method is it may require measuring the long-time behaviour of the time evolution of the lepton family number.
Such a measurement may prove to be experimentally difficult.
In the next section, we propose different quantities that are directly related to the Majorana type-phases.
These quantities are from the derivatives of the expectation values at a slice in time.
%%%%%%%%%%%%%%%%%%%%%%%%%%%%%%%%%%%%%%%%
\section{Determination of Majorana type-phases and the lightest neutrino masses with the lepton numbers}\label{sec:determineMajoranatypephasesandlightestmass}
In this section, we derive a formula to determine the lightest neutrino mass and the Majorana phases from the derivatives of the expectation values at a slice in time.
Constraining the lightest neutrino mass and the Majorana phases has already been discussed in the context of neutrinoless double $\beta$ decays~\cite{Pascoli:2002qm,BhupalDev:2013ntw}.
We consider two similar setups,
\begin{enumerate}
    \item there are three massive neutrinos where two Majorana phases are allowed,
    \item there is a massless neutrino and only one Majorana phase is allowed.
\end{enumerate}
To begin, we take the second-order time derivative of Eq.(\ref{eq:lepton_M}) at the initial time $t=0$,
\begin{equation}
  \begin{split}
    \frac{d^2}{dt^2}\bra*{\sigma}L_\alpha^M(t)\ket*{\sigma}\vert_{t=0} &= -\sum_{i,j}^3 \Re(U_{\alpha i}^\ast U_{\sigma i} U_{\alpha j} U_{\sigma j}^\ast)\left( m_i^2+m_j^2 \right)- 2\sum_{i,j}^3 {\rm Re}(U_{\alpha i}^\ast U_{\sigma i}^\ast U_{\alpha j} U_{\sigma j}) m_i m_j
\\
    &=-2\sum_{i}^3 \delta_{\alpha \sigma}\abs{U_{\sigma i}}^2 m_i^2-2\sum_{i,j}^3 \Re(U_{\alpha i}^\ast U_{\sigma i}^\ast U_{\alpha j} U_{\sigma j}) m_i m_j.
  \label{eq:derivetiveslepton_M}
  \end{split}
\end{equation}
Similar to what we found in the previous section, the first term is independent of the Majorana phases and the second term depends on the Majorana phases though the PMNS combination $\Re(U^\ast_{\alpha i} U^\ast_{\sigma i} U^{}_{\alpha j} U^{}_{\sigma j})$.
Then, the second-order derivative of the total lepton number $L^M(t)=\sum_{\alpha} L_\alpha^M(t)$ is given by,
\begin{equation}
    \frac{d^2}{dt^2}\bra*{\sigma} L^M(t)\ket*{\sigma}|_{t=0}  = -4 \sum_{i}^3   m_i^2 |U_{\sigma i}|^2.
 \label{eq:derivetivesTotallepton_M}
\end{equation}
This is similar to the first term in the result of Eq.(\ref{eq:derivetiveslepton_M}).
We use that result for the total lepton number of Eq.(\ref{eq:derivetivesTotallepton_M}) to rewrite the first term of Eq.(\ref{eq:derivetiveslepton_M}) leading to,
\begin{equation}
    \frac{d^2}{dt^2}\bra*{\sigma}L_\alpha^M(t)\ket*{\sigma}|_{t=0} =\delta_{\alpha \sigma}\frac{1}{2}
      \frac{d^2}{dt^2}\bra*{\sigma} L^M(t)\ket*{\sigma}|_{t=0}
     -2\sum_{i,j}^3 {\rm Re}(U_{\alpha i}^* U_{\sigma i}^* U_{\alpha j} U_{\sigma j}) m_i m_j.
\label{eq:2ndderi}
\end{equation}
Using Eq.(\ref{eq:derivetivesTotallepton_M}), we also derive equations for the lightest neutrino mass, normal hierarchy and inverted hierarchy cases, in terms of the second-order derivative of the total lepton number.
\begin{align}
    m_1^2=- \frac{1}{4} \frac{d^2}{dt^2}\bra*{\sigma} L^M(t)\ket*{\sigma}|_{t=0}
-\Delta m^2_{21}|U_{\sigma 2}|^2-\Delta m^2_{31} |U_{\sigma 3}|^2 &&& \text{ Normal,}
\label{eq:lightestneutrinomassN}
\\
    m_3^2=-\frac{1}{4} \frac{d^2}{dt^2}\bra*{\sigma} L^M(t)\ket*{\sigma}|_{t=0}
-\Delta m^2_{13}|U_{\sigma 1}|^2-\Delta m^2_{23}|U_{\sigma 2}|^2 &&& \text{ Inverted.}
\label{eq:lightestneutrinomassI}
\end{align}

To investigate the Majorana type-phases, we consider two cases for the lepton families.
First we set $\alpha=\sigma$ in Eq.(\ref{eq:2ndderi}) and obtain the following formula,
\begin{equation}
    \frac{d^2}{dt^2}\bra*{\sigma}L_\sigma^M(t)\ket*{\sigma}|_{t=0}
    = \frac{1}{2} \frac{d^2}{dt^2}\bra*{\sigma} L^M(t)  \ket*{\sigma}|_{t=0}
    -2\sum_{i,j}^3 (U_{\sigma i}^*U_{\sigma j})^2 m_i m_j.
    \label{eq:timederivative1}
\end{equation}
Then we set $\alpha \ne \sigma$ in Eq.(\ref{eq:2ndderi}) to obtain,
\begin{equation}
    \frac{d^2}{dt^2}\bra*{\sigma}L_\alpha^M(t)\ket*{\sigma}|_{t=0} =
     -2\sum_{i,j}^3 {\rm Re}(U_{\alpha i}^* U_{\alpha j}U_{\sigma i}^* U_{\sigma j}) m_i m_j.
\label{eq:timederivative2} 
\end{equation}
Next, by specifying the family indices $\sigma$ and $\alpha$ in Eq.(\ref{eq:timederivative1}) and Eq.(\ref{eq:timederivative2}) we can clarify the dependencies on the Majorana type-phases from Eqs.(\ref{eq:majoranaTypeBeta}-\ref{eq:majoranaTypeGamma}).
We take $\sigma=e$ in Eq.(\ref{eq:timederivative1}) to show the dependencies on the Majorana type-phases $\beta_1$ and $\gamma_1$,
\begin{multline}
    \frac{d^2}{dt^2}\bra*{e}L_e^M(t)\ket*{e}\vert_{t=0}=\frac{1}{2}\frac{d^2}{dt^2}\bra*{e} L^M(t) \ket*{e}\vert_{t=0} -2 \sum_{i=1}^3 m_i^2 \abs{U_{ei}}^4 -4 m_1 m_2 \abs{U_{e1} U^\ast_{e2}}^2\cos(2\beta_1)
\\
    -4 m_2 m_3 \abs{U_{e2} U^\ast_{e3}}^2 \cos(2\beta_1-2\gamma_1) -4 m_3 m_1 \abs{U_{e1} U^\ast_{e3}}^2 \cos(2\gamma_1).
  \label{eq:timederivative3}
\end{multline}
In contrast, for Eq.(\ref{eq:timederivative2}) we take $\sigma=e$ and $\alpha=\mu$,
\begin{multline}
    \frac{d^2}{dt^2}\bra*{e}L_\mu^M(t)\ket*{e}\vert_{t=0} = -2 \sum_{i}^3  \abs{U_{\mu i}}^2 \abs{U_{e i}}^2 m_i^2 -4 \abs{U_{\mu 1}^\ast U_{\mu 2}U_{e 1}^\ast U_{e 2}}\cos(\beta_1+\beta_2)  m_1 m_2
\\
    \qquad-4 \abs{U_{\mu 2}^\ast U_{\mu 3}U_{e 2}^\ast U_{e 3}} \cos(\beta_1-\gamma_1+\beta_2-\gamma_2) m_2 m_3 
\\
    -4 \abs{U_{\mu 3}^\ast U_{\mu 1}U_{e 3}^\ast U_{e 1}}\cos(\gamma_1+\gamma_2) m_3 m_1,
\end{multline}
where we have dependence on the additional Majorana type-phases $\beta_2$ and $\gamma_2$ compared to Eq.(\ref{eq:timederivative3}).
Lastly we take $\alpha = \tau$,
\begin{multline}\label{eq:timederivative4} 
    \frac{d^2}{dt^2}\bra*{e}L_\tau^M(t)\ket*{e}|_{t=0} = -2 \sum_{i}^3  \abs{U_{\tau i}}^2 \abs{U_{e i}}^2  m_i^2 -4 \abs{U_{\tau 1}^\ast U_{\tau 2}U_{e 1}^\ast U_{e 2}}\cos(\beta_1+\beta_3)  m_1 m_2
\\
    \qquad-4 \abs{U_{\tau 2}^\ast U_{\tau 3}U_{e 2}^\ast U_{e 3}} \cos(\beta_1-\gamma_1+\beta_3-\gamma_3) m_2 m_3
\\
    -4 \abs{U_{\tau 3}^\ast U_{\tau 1}U_{e 3}^\ast U_{e 1}}\cos(\gamma_1+\gamma_3) m_3 m_1,
\end{multline}
which results in the dependence on the Majorana type-phases $\beta_3$ and $\gamma_3$.

As discussed near the end of section \ref{sec:MajoranaTypePhases}, the Majorana type-phases $\beta_i$ and $\gamma_i$ change for variation of the Majorana phases $\Delta \alpha_{21}$ and $\Delta \alpha_{31}$,
\begin{gather}\label{eq:Betamajoranaphasevar}
    \beta_i +\beta_j \to \beta_i+\beta_j-\Delta \alpha_{21},
\\ \label{eq:Gammamajoranaphasevar}
    \gamma_i+\gamma_j \to \gamma_i+\gamma_j-\Delta \alpha_{31}. 
\end{gather}
The time derivatives of all the three lepton family numbers in Eq.(\ref{eq:timederivative3}-\ref{eq:timederivative4}) are dependent on summations of the Majorana type-phases i.e., $\beta_i +\beta_j$ and $\gamma_i +\gamma_j$.
From Eqs.(\ref{eq:Betamajoranaphasevar}-\ref{eq:Gammamajoranaphasevar}) we know $\beta_i +\beta_j$ and $\gamma_i +\gamma_j$ are sensitive to the Majorana phases $\Delta \alpha_{21}$ and $\Delta \alpha_{31}$.
Thus, the two Majorana phases $\alpha_{21}$ and $\alpha_{31}$ can be determined by knowing the time derivatives of the two or more lepton family numbers.
In addition, we can resolve the lightest neutrino mass from the time derivatives of the total lepton number in Eqs.(\ref{eq:lightestneutrinomassN}-\ref{eq:lightestneutrinomassI}).
This is a method to determine the triangle orientations and the lightest neutrino mass based
on lepton number oscillations.

%%%%%%%%%%%%%%%%%%%%%%%%%%%%%%%%%%%%%%%%
\section{Considerations when the lightest neutrino is massless}\label{sec:lightestismassless}
In this section, we consider the lightest neutrino to be massless.
An example of such a setup is the $(3,2)$ Type-I seesaw model, which can predict a massless neutrino from three active neutrinos and two heavy right-handed Majorana neutrinos~\cite{Ma:1998zg}.
In this framework, which active neutrino is massless depends on the hierarchy,
\begin{align} \label{eq:normalhiermassless}
    m_1=0, && 0 < m_2 < m_3, && \text{Normal hierarchy,}
\\ \label{eq:inverthiermassless}
    m_3=0, && 0 < m_1 < m_2, && \text{Inverted hierarchy.}
\end{align}
Because a massless neutrino is free to be re-phased, a few of the six Majorana type-phases from the three massive Majorana neutrinos in the previous section~\ref{sec:MajoranaTypePhases} are no longer invariants.

In the normal hierarchy case, the invariant combinations of the three $\beta_i$'s and three $\gamma_i$'s become,
\begin{align}\label{MajoranaTypeNormal}
    \arg(U_{e2} U_{e3}^\ast)=\gamma_1-\beta_1, && \arg(U_{\mu2} U_{\mu3}^\ast)=\gamma_2-\beta_2, && \arg(U_{\tau2} U_{\tau 3}^\ast)=\gamma_3-\beta_3.
\end{align}
Consequently, only the orientation of the triangle 3 in Fig.~\ref{fig:tripletriangles} is invariant and a physical quantity.
This is because, $\beta_i$ and $\gamma_i$ are no longer re-phasing invariants by themselves.
Instead, $\beta_i$ and $\gamma_i$ can also be formed into four invariant quartets,
\begin{gather}\label{eq:quartetBeta21}
    \arg(U_{e1}^\ast U_{e2}U_{\mu1} U_{\mu2}^\ast)=\beta_2-\beta_1,
\\ \label{eq:quartetBeta31}
    \arg(U_{e1}^\ast U_{e2}U_{\tau1} U_{\tau 2}^\ast)=\beta_3-\beta_1,
\\ \label{eq:quartetGamma21}
    \arg(U_{e1}^\ast U_{e3}U_{\mu1} U_{\mu3}^\ast)=\gamma_2-\gamma_1,
\\ \label{eq:quartetGamma31}
    \arg(U_{e1}^\ast U_{e3}U_{\tau1} U_{\tau 3}^\ast)=\gamma_3-\gamma_1.
\end{gather}
Those four arguments defined in Eqs.(\ref{eq:quartetBeta21}-\ref{eq:quartetGamma31}) and the Majorana type-phases of Eq.(\ref{MajoranaTypeNormal}) are not independent.
As an example, if we choose the first Majorana type-phase $\gamma_1-\beta_1$ in Eq.(\ref{MajoranaTypeNormal}) as an invariant, and also choose the four arguments of the quartets as independent re-phasing invariants.
The other two Majorana type-phases can be written using them,
\begin{equation}
    \begin{split}
        \gamma_i-\beta_i&=(\gamma_1-\beta_1)+(\gamma_i-\beta_i)-(\gamma_1-\beta_1)
    \\
        &=(\gamma_1-\beta_1)+(\gamma_i-\gamma_1)-(\beta_i-\beta_1) \qquad (i=2,3).
    \end{split}
\label{eq:relMt}
\end{equation}

A similar situation occurs for the inverted hierarchy case, with the massless neutrino $m_3=0$.
The re-phasing invariant Majorana type-phases become $\beta_i$, for $i=1,2,3$, and the orientation of triangle 1 in Fig.~\ref{fig:tripletriangles} is invariant.
Then, one can choose a Majorana type-phase $\beta_1$, the two invariants of Eq.(\ref{eq:quartetGamma21}) and Eq.(\ref{eq:quartetGamma31}), and two extra arguments,
\begin{gather}\label{eq:quartetBeta-Gamma12}
    \arg(U_{e3}^\ast U_{e2}U_{\mu3} U_{\mu2}^\ast)=(\gamma_1-\beta_1)-(\gamma_2-\beta_2),
\\ \label{eq:quartetBeta-Gamma13}
    \arg(U_{e3}^\ast U_{e2}U_{\tau3} U_{\tau 2}^\ast)=(\gamma_1-\beta_1)-(\gamma_3-\beta_3).
\end{gather}
As a result, the remaining $\beta_2$ and $\beta_3$ can be written as linear combinations of $\beta_1$ and the four invariant quartets in Eqs.(\ref{eq:quartetGamma21}, \ref{eq:quartetGamma31}, \ref{eq:quartetBeta-Gamma12}, \ref{eq:quartetBeta-Gamma13}),
\begin{gather}
  \beta_2=(\gamma_1-\beta_1)-(\gamma_2-\beta_2)+(\gamma_2-\gamma_1) +\beta_1,
  \\
  \beta_3=(\gamma_1-\beta_1)-(\gamma_3-\beta_3)+(\gamma_3-\gamma_1) +\beta_1.
\end{gather}
This shows how a few of the six Majorana type-phases are not invariants for situations with a massless neutrino.

%%%%%%%%%%%%%%%%%%%%%%%%%%%
\subsection{Neutrinoless double beta decay and lepton number}
In this subsection, we relate one independent Majorana type-phase to physical observables, such as $\abs{m_{\nu ee}}$.
This pertains to the neutrinoless double beta decay rate and the time evolution of lepton numbers.
For the normal hierarchy case, the $\abs{m_{\nu ee}}$ is given by the following formula,
\begin{equation}\label{eq:doubleBetanormalmass}
    \abs{m_{\nu e e}}^2_\text{norm} = m_2^2 \abs{U_{e 2}}^4 +m_3^2 \abs{U_{e 3}}^4 + 2 m_2 m_3 \abs{U_{e 2}}^2 \abs{U_{e 3}}^2 \cos{2 (\gamma_1-\beta_1)}.
\end{equation}
Whereas for the inverted hierarchy case it is,
\begin{equation}\label{eq:doubleBetainvertedmass}
    \abs{m_{\nu e e}}^2_\text{inv} = m_1^2 \abs{U_{e 1}}^4 +m_2^2 \abs{U_{e 2}}^4 + 2 m_1 m_2 \abs{U_{e 1}}^2 \abs{U_{e 2}}^2 \cos{2\beta_1}.
\end{equation}
From the expressions in Eq.(\ref{eq:doubleBetanormalmass}-\ref{eq:doubleBetainvertedmass}), the determination of $\abs{m_{\nu e e}}_{\text{norm},\text{inv}}$ is sufficient to identify a single Majorana type-phase; provided the other moduli of the PMNS matrix elements in Eq.(\ref{eq:DiracPMNSmatrix}) and two non-zero masses of the neutrinos are extracted from the neutrino oscillation experiments.

Next, we investigate the time evolution of lepton number when the lightest neutrino is massless, as in Eqs.(\ref{eq:normalhiermassless}-\ref{eq:inverthiermassless}).
From Eq.(\ref{eq:lightestneutrinomassN}), for the normal hierarchy case, the second-order time derivative of the total lepton number is related to the combination of the two non-vanishing masses;
\begin{equation}
    \frac{1}{4} \frac{d^2}{dt^2}\bra*{\sigma} L^M(t)\ket*{\sigma}|_{t=0} = - m^2_{2}\abs{U_{\sigma 2}}^2- m^2_{3} \abs{U_{\sigma 3}}^2.
  \label{eq:m1=0}
\end{equation}
Then we can rewrite the second-order time derivatives of the electron, muon, and tauon lepton numbers of Eqs.(\ref{eq:timederivative3}-\ref{eq:timederivative4}) as,
\begin{gather}\label{eq:timederivativeNormalmassless1}
    \frac{d^2}{dt^2}\bra*{e} L_e^M(t)\ket*{e}\vert_{t=0} = \frac{d^2}{dt^2}\bra*{e}L^M(t) \ket*{e}\vert_{t=0}-4 m_2 m_3 \abs{U_{e2} U^\ast_{e3}}^2 \cos(2\beta_1-2\gamma_1),
\\ \label{eq:timederivativeNormalmassless2}
    \frac{d^2}{dt^2}\bra*{e} L_\mu^M(t)\ket*{e}\vert_{t=0} =-2 \sum_{i=2}^3  \abs{U_{\mu i}}^2 \abs{U_{e i}}^2  m_i^2 -4 \abs{U_{\mu 2}^\ast U_{\mu 3}U_{e 2}^\ast U_{e 3}} \cos(\beta_1-\gamma_1+\beta_2-\gamma_2) m_2 m_3,
\\ \label{eq:timederivativeNormalmassless3}
    \frac{d^2}{dt^2}\bra*{e} L_\tau^M(t)\ket*{e}\vert_{t=0} =-2 \sum_{i=2}^3 \abs{U_{\tau i}}^2 \abs{U_{e i}}^2 m_i^2 -4 \abs{U_{\tau 2}^* U_{\tau 3}U_{e 2}^* U_{e 3}} \cos(\beta_1-\gamma_1+\beta_3-\gamma_3) m_2 m_3.
\end{gather}
The second-order time derivative for the normal hierarchy in Eq.(\ref{eq:timederivativeNormalmassless1}) can be used to determine one of the Majorana like-phases; $\gamma_1-\beta_1$.
The other two Majorana type-phases, $\gamma_i-\beta_i$ $(i=2,3)$, can be determined through Eq.(\ref{eq:relMt}) by knowing the arguments of the quartets in Eqs.(\ref{eq:quartetBeta21})-(\ref{eq:quartetGamma31}). 
This allows us to learn of the orientation of  triangle 3 in Fig.\ref{fig:tripletriangles}.
Alternatively, with the three second-order time derivatives of Eqs.(\ref{eq:timederivativeNormalmassless1})-(\ref{eq:timederivativeNormalmassless3}) one can determine three Majorana type-phases $\gamma_i-\beta_i (i=1-3)$.

In the inverted hierarchy case $m_3$ is the lightest neutrino mass and the second-order time derivative of the total lepton number becomes,
\begin{equation}
    \frac{1}{4} \frac{d^2}{dt^2}\bra*{\sigma} L^M(t)\ket*{\sigma}\vert_{t=0}= -m^2_{2}\abs{U_{\sigma 2}}^2- m^2_{1}\abs{U_{\sigma 1}}^2.
  \label{eq:lm3=0} 
\end{equation}
The major difference between the normal hierarchy of Eq.(\ref{eq:m1=0}) and the inverted hierarchy of Eq.(\ref{eq:lm3=0}) is the last mass multiplied by PMNS matrix term.
This modifies the second-order time derivatives to become,
\begin{gather}\label{eq:timederivativeInverted1}
    \frac{d^2}{dt^2}\bra*{e}L_e^M(t)\ket*{e}\vert_{t=0}=\frac{d^2}{dt^2}\bra*{e} L^M(t) \ket*{e}\vert_{t=0} -4 m_1 m_2 \abs{U_{e1} U^\ast_{e2}}^2\cos(2\beta_1),
\\ \label{eq:timederivativeInverted2}
    \frac{d^2}{dt^2}\bra*{e}L_\mu^M(t)\ket*{e}\vert_{t=0} = -2 \sum_{i}^2  \abs{U_{\mu i}}^2 \abs{U_{e i}}^2 m_i^2 -4 \abs{U_{\mu 1}^\ast U_{\mu 2}U_{e 1}^\ast U_{e 2}}\cos(\beta_1+\beta_2)  m_1 m_2,
\\ \label{eq:timederivativeInverted3}
    \frac{d^2}{dt^2}\bra*{e}L_\tau^M(t)\ket*{e}|_{t=0} = -2 \sum_{i}^2  \abs{U_{\tau i}}^2 \abs{U_{e i}}^2  m_i^2 -4 \abs{U_{\tau 1}^\ast U_{\tau 2}U_{e 1}^\ast U_{e 2}}\cos(\beta_1+\beta_3)  m_1 m_2.
\end{gather}
Similar to the normal hierarchy equations of Eqs.(\ref{eq:timederivativeNormalmassless1}-\ref{eq:timederivativeNormalmassless3}), the second-order time derivatives of the inverted hierarchy in Eqs.(\ref{eq:timederivativeInverted1}-\ref{eq:timederivativeInverted3})
allow for a complete determination of the Majorana type-phases $\beta_i$
and the orientation of triangle 1 in  figure \ref{fig:tripletriangles}.

%%%%%%%%%%%%%%%%%%%%%%%%%%%%%%%%%%%%%%%%%%%%%%%%%%%%%%%
\subsection{Numerical illustration for two neutrino generations}
In this subsection, we illustrate the relations among the lightest neutrino mass, a Majorana type-phase, and the second-order time derivative of lepton numbers. 
Because a realistic model with three generations of neutrinos is more challenging to analyze, we focus on the toy model of two generations of active neutrinos.
In this model, there are only two Majorana type-phases, $\beta_1$ and $\beta_2$ defined in Eq.(\ref{eq:majoranaTypeBeta}).
They are not independent because the unitary relation
\begin{equation}
  U_{e1}U_{e2}^\ast=-U_{\mu1}U_{\mu 2}^\ast, 
\end{equation}
holds.
This leads to the Majorana type-phases being related by a phase shift,
\begin{equation}
  \beta_1=\arg(U_{e1}U_{e2}^\ast)=\arg(U_{\mu1}U_{\mu 2}^\ast)-\pi=\beta_2-\pi.
\end{equation}
Adopting the following parametrization for a 2 by 2 unitary PMNS matrix,
\begin{equation}
  \begin{pmatrix}
    \cos \theta_{12} & \sin \theta_{12} e^{i \frac{\alpha_{21}}{2}} \\
    -\sin \theta_{12} & \cos \theta_{12} e^{i \frac{\alpha_{21}}{2}} \\
  \end{pmatrix},
  \label{eq:2by2PMNS}
\end{equation}
the Majorana type-phase $\beta_1$ is related to a Majorana phase as  $\beta_1=-\frac{\alpha_{21}}{2}$.
Contrary to the three generation model, an equation similar to Eq.(\ref{eq:theta12}) for $\tan{\theta_{12}}$ can not be written in terms of the Majorana type-phase. 

In this model, we assume that the mass squared difference $\Delta m^2_{21}$ and the mixing angle $\theta_{12}$ can be measured by oscillation experiments.
Then, the remaining parameters are the lightest neutrino mass $m_1$ and the Majorana type-phase $\beta_1$.
Assuming the expectation values of lepton numbers are measurable, $m_1$ and $\beta_1$ could be determined by the time evolution of lepton numbers.
To demonstrate this, we write the time evolution of the electron and muon numbers for the two generation model,
\begin{align} \label{eq:L_nu e} 
    \begin{split}
        \bra{e}L_e(t)\ket{e} = {} & c_{12}^4 \left( 1 - \frac{2m_1^2 \sin^2 (E_1 t)}{E_1^2} \right) + s_{12}^4 \left( 1 - \frac{2 m_2^2\sin^2 (E_2 t)}{E_2^2} \right)
    \\
        & + s_{12}^2 c_{12}^2 \left\{ \left( 1 + \frac{q^2 - m_1 m_2 \cos(2 \beta_1 )}{E_1 E_2} \right) \cos \{ (E_1 - E_2) t \} \right.
    \\
        & \left. + \left( 1 -\frac{q^2 - m_1 m_2 \cos (2\beta_1 )}{E_1 E_2} \right) \cos \{ (E_1 + E_2)t \} \right\},
    \end{split}
\\  \label{eq:L_nu mu}
    \begin{split}
        \bra{e}L_{\mu}(t)\ket{e} = {} & c_{12}^2 s_{12}^2 \left( \left( 1 - \frac{2 m_1^2 \sin^2 (E_1 t)}{E_1^2} \right) + \left( 1 - \frac{2m_2^2 \sin^2 (E_2 t)}{E_2^2} \right) \right)
    \\
        & - s_{12}^2 c_{12}^2 \left\{ \left( 1 + \frac{q^2 - m_1 m_2 \cos(2 \beta_1)}{E_1 E_2} \right) \cos \{ (E_1 - E_2) t \} \right.
    \\
        & \left. + \left( 1 - \frac{q^2 - m_1 m_2 \cos (2 \beta_1)}{E_1 E_2} \right) \cos \{ (E_1 + E_2)t \} \right\} .
    \end{split}
\end{align}
Similar to the three generation model, the expectation values of the total lepton number $L(t)=L_e(t)+L_\mu(t)$ and the difference $L_{e-\mu}(t)=L_e(t)-L_\mu(t)$ can be computed.

\begin{figure}[htp]
    \begin{center}
    \includegraphics[width=0.5\textwidth]{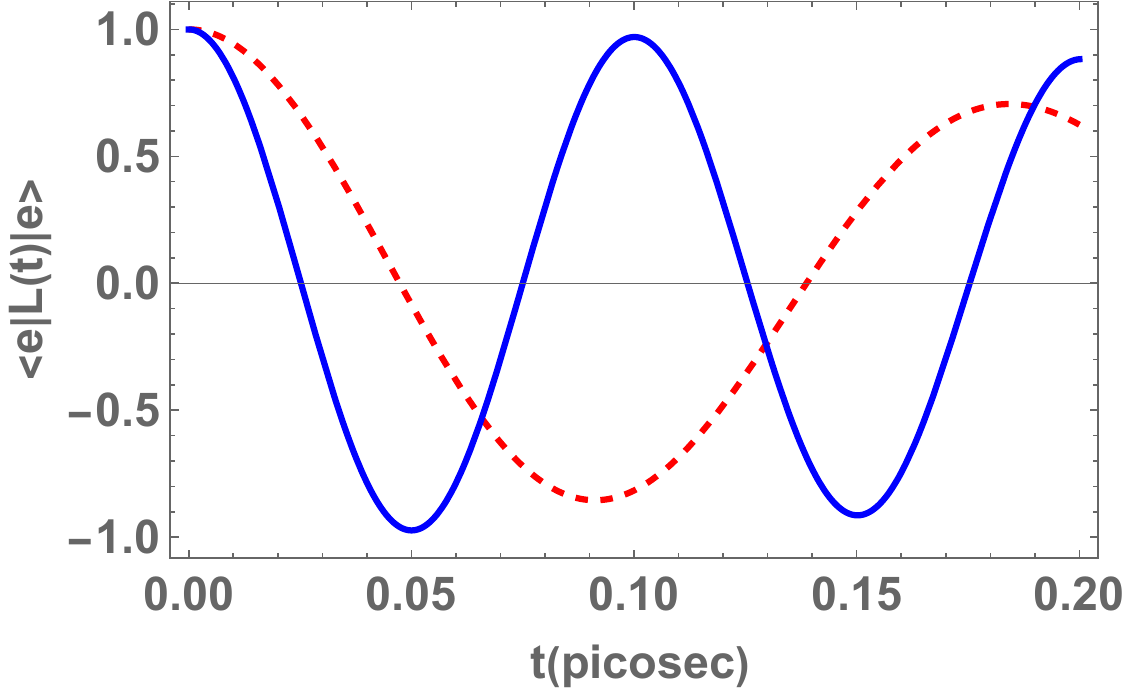}
    \caption{\label{fig:eplusmu} The time dependence of the total lepton number for different lightest neutrino masses.  The momentum  $q=0.002$(eV). 
    The dashed line shows the case for $m_1=0.01$(eV) and the solid line shows the case for $m_1=0.02$(eV).  We use $\Delta m^2_{21}=7.42 \times 10^{-5}\text{eV}^2$ and $\sin(\theta_{12})=0.551$. }
    \end{center}
\end{figure}

Plotting $\bra{e}L(t)\ket{e}$ in Fig.~\ref{fig:eplusmu} shows the evolution of the total lepton number depends on the lightest neutrino mass $m_1$.
At $t=0$ the larger $m_1=0.02$ eV features a sharper decrease than the case of $m_1 = 0.01$ eV. 
This difference in behavior can be also understood from a straightforward calculation using $L(t)$ and Eqs.(\ref{eq:L_nu e}-\ref{eq:L_nu mu}).
The second order time derivative for the total lepton number at $t=0$ is given by,
\begin{equation} \label{eq:5.5}
    \frac{d^2}{dt^2} \bra{e}L(t)\ket{e} \Bigr|_{t = 0} = - 4 (m_1^2 c_{12}^2 + m_2^2 s_{12}^2).
\end{equation}
For a fixed angle and mass squared difference, a larger $m_1$ results in a faster change in the slope of $\bra{e}L(t)\ket{e}$.
Using that insight, the lightest neutrino mass can be determined with the second derivative of Eq.(\ref{eq:5.5}), the mixing angle, and the mass squared difference,
\begin{equation}
  m_1^2  = - \frac{1}{4}\frac{d^2}{dt^2} \bra{e}L(t)\ket{e} \Bigr|_{t = 0} - \Delta m^2_{21} s_{12}^2.
  \label{eq:m1}
\end{equation}

\begin{figure}[htp]
    \begin{center}
    \includegraphics[width=0.5\textwidth]{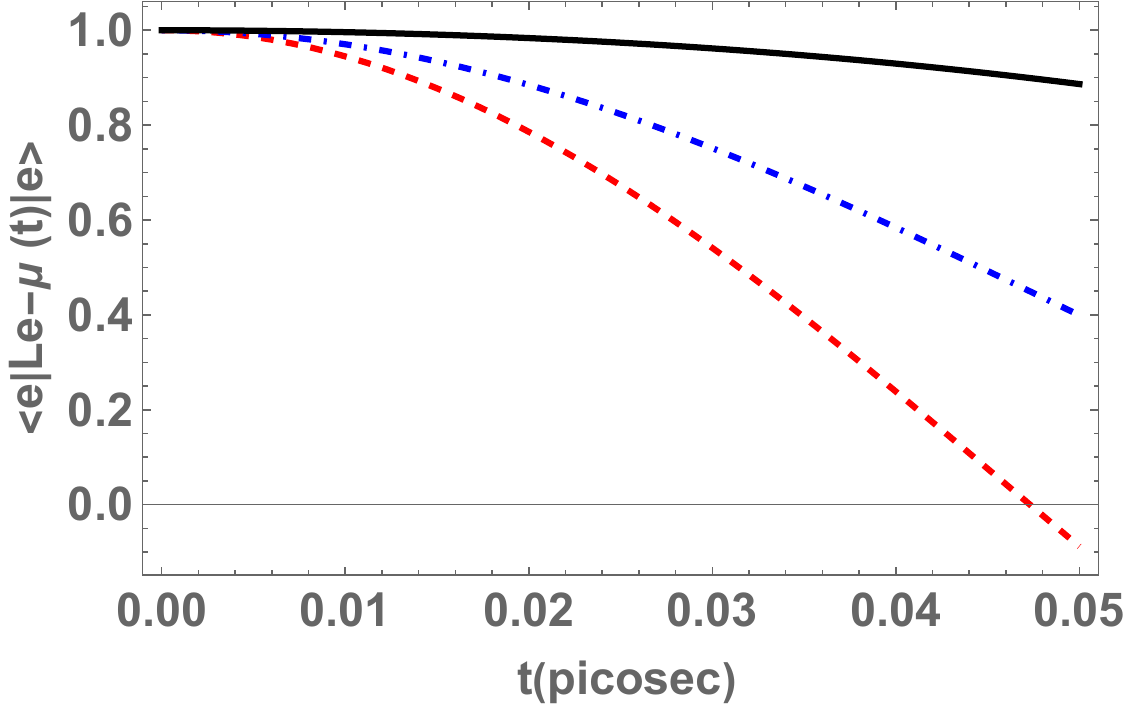}
    \caption{\label{fig:eminusmu} The time dependence of $L_{e-\mu}=L_e-L_\mu $for different choice of the Majorana type-phase $\beta_1$. $q=0.002$(eV) $m_1=0.01$(eV) and $\sin (\theta_{12})=0.551$.
    The dashed line shows the case for $\beta_1=0$.
    The dot-dashed line shows the case for $2\beta_1=\frac{\pi}{2}$ and solid line shows the case for $2\beta_1=\pi$.
$\Delta m^2_{21}$ is the same as in Fig.4.}
 \end{center}
\end{figure}

The second order time derivative for the electron minus muon number $L_{e-\mu}$ at $t=0$ is given as follows,
\begin{equation} \label{eq:5.6}
    \frac{d^2}{dt^2}  \bra{e}L_{e-\mu}(t)\ket{e} \Bigr|_{t = 0} = - 4 \abs{m_{\nu ee}}^2 _\text{two gene.},
\end{equation}
where $\abs{m_{\nu ee}}^2 _\text{two gene.}$ is similar to Eq.(\ref{eq:doubleBetainvertedmass}) for the inverted hierarchy case with $m_3=0$.
By substituting the mixing angle for $|U_{ei}|$ ($i=1,2$) from Eq.(\ref{eq:2by2PMNS}) the explicit form of $\abs{m_{\nu ee}}^2 _\text{two gene.}$ is, 
\begin{equation}\label{2genemee}
    \abs{m_{\nu e e}}^2_\text{two gene.} = m_1^2 {c_{12}}^4 +m_2^2 {s_{12}}^4 + 2 m_1 m_2 {c_{12}}^2 {s_{12}}^2 \cos{2\beta_1}.
\end{equation}
Eq.(\ref{eq:5.6}) tells us that $\bra{e}L_{e-\mu}(t)\ket{e}$ has a sharper change in slope for larger $\abs{m_{\nu ee}}_\text{two gene.}$.
This is demonstrated in Fig.(\ref{fig:eminusmu}), which shows the dependence of $\bra{e}L_{e-\mu}(t)\ket{e}$ on the Majorana type-phase $\beta_1$.
Because, as shown in Eq.(\ref{2genemee}), $\abs{m_{\nu ee}}_\text{two gene.}$ is the largest for $\beta_1=0$ and the smallest for $\beta_1=\pi$, Fig.(\ref{fig:eminusmu}) numerically confirms the dependence of $\abs{m_{\nu ee}}_\text{two gene.}$ in Eq.(\ref{eq:5.6}).
Using that understanding, the Majorana type-phase $\beta_1$ can be determined from the second order derivative of Eqs.(\ref{eq:5.6}), the mixing angle, and the mass squared difference.
\begin{equation}
  \cos (2\beta_1)  = \frac{-\frac{1}{4}\frac{d^2}{dt^2}\bra{e}L_{e-\mu}(t)\ket{e} \bigr|_{t = 0} - (m_1^2 c_{12}^4 + m_2^2 s_{12}^4)}{2 m_1 m_2 c_{12}^2 s_{12}^2}.
  \label{eq:beta1}
\end{equation}
Combined, Eqs.(\ref{eq:m1}-\ref{eq:beta1}) can be used to determine the lightest neutrino mass and the Majorana type-phase once the time derivatives for $\ddot{L}|_{t=0}$ and $\ddot{L}_{e-\mu}|_{t=0}$ are measured.

%%%%%%%%%%%%%%%%%%%%%%%%%%%%%%%%%%%%%%%%%%%%%%%%%
\section{Neutrino masses, the Majorana phase, and the mixing angle in terms of elements from the effective Majorana mass matrix}\label{sec:Meff}
At the beggining of this section, we introduce a parametrization of the effective Majorana mass matrix $m_\nu$ in the charged lepton flavor basis for $N$ generation lepton families.
We parametrize the $N \times N $ complex symmetric matrix $m_\nu$ with invariants under the possible rephasing for charged leptons families in Eq.(\ref{weaktr}) below.
This transformation removes the phases of the diagonal elements of the effective Majorana mass matrix $m_\nu$ while keeping the charged lepton mass matrix real diagonal.
This parametrization includes $\frac{N(N+1)}{2}$ real parts corresponding to the moduli of the $N \times N$ symmetric matrix.
It also includes $\frac{N(N-1)}{2}$ invariant phases of the off-diagonal elements of the mass matrix.
If the rank of the mass matrix is $N$ and the eigenvalues are not degenerate, these parameters determine the $N$ non-zero masses of neutrinos and $\frac{N(N-1)}{2}$ mixing angles of PMNS matrix through the diagonalization of the mass matrix. 
The $\frac{N(N-1)}{2}$ imaginary parameters are related to the CP violating phases in PMNS matrix.
The CP violating phases consist of $N-1$ Majorana phases and $\frac{(N-2)(N-1)}{2}$ Dirac Kobayashi-Maskawa phases. 

We start with the effective Majorana mass matrix $m_\nu$ in a basis where charged lepton mass matrix $m_a$ is real and diagonal. Hereafter, the Latin letters denote the charged lepton families, $a=e,\mu,\tau,\dotsc$ and the Lagrangian is written as,
\begin{equation}
  \begin{split}
    \mathcal{L}_{lepton} & = -\frac{1}{2}  \overline{(\nu_{a L})^c} ({m_{\nu}})_{ab} \nu_{b L}-h.c. -\overline{l_a} m_a l_a
  \\
    & =-\frac{1}{2} \overline{(\nu^\prime_{a L})^c} ({m^0_{\nu}})_{ab} \nu^\prime_{b L}-h.c. -\overline{l^\prime_a} m_a l^\prime_a,
  \end{split}
\end{equation}
where ${(m^0_{\nu})}_{ab}$ is the effective Majorana mass matrix obtained after a rephasing of the lepton doublets $L_a=\begin{pmatrix} \nu_{a L} & l_{a L} \end{pmatrix}^T$ and the SU(2) singlets $l_{R a}$,
\begin{align}
  L_a=e^{i \delta_a} L^\prime_a, && l_{R a}=e^{i \delta_a} l^\prime_{R a}.
\label{weaktr}
\end{align}
Under that transformation, the electroweak interaction does not change whereas $({m_{\nu}})_{ab}$ is modifed to be,
\begin{equation}
  ({m^0_{\nu}})_{ab} = ({m_{\nu}})_{ab} e^{i  (\delta_a+\delta_b)}.
  \label{m0}
\end{equation}
This implies the arguments of $m_\nu$ and $m^0_\nu$ are related to each other as,
\begin{align}
  \arg ({m_{\nu}})_{ab}\equiv \theta_{ab}, && \arg({m^{0}_{\nu}})_{ab}\equiv \theta^\prime_{ab}=\theta_{ab}+\delta_a+\delta_b.
\end{align}
One can freely choose $\delta_a$ such that the phases of the diagonal elements of $m^0_{\nu}$ vanish,
\begin{equation}
  \theta^\prime_{aa}=0.
\end{equation}
Then $\delta_a$ can be found from, 
\begin{align}
  \theta_{aa}+2\delta_a=0, && \delta_a=-\frac{\theta_{aa}}{2}.
\end{align}
Next, the non-vanishing $\frac{N(N-1)}{2}$ phases in $({m^{0}_{\nu}})_{ab}$ for $a \ne b$ are determined to be,
\begin{equation}
  \arg({m^{0}_{\nu}})_{ab}=\theta^\prime_{ab}=\theta_{ab}-\frac{\theta_{aa}+\theta_{bb}}{2}.
\label{eq:argm0eff}
\end{equation}
The phases $\theta^\prime_{ab}$ are invariant under the change of the weak basis as given by Eq.(\ref{weaktr}).
The above consideration leads to the conclusion that one can start with the following effective Majorana matrix without loss of generality,
\begin{equation}
  {m^{0}_{\nu}}_{ab}=\delta_{ab} |m_{aa}|+(1-\delta_{ab}) |m_{ab}|e^{i\theta^\prime_{ab}},
\label{para}
\end{equation}
where we denote ${m^{0}_{\nu}}_{aa}\equiv |m_{aa}|$ for the diagonal elements and ${m^{0}_{\nu}}_{ab}\equiv m_{ab}=|m_{ab}| e^{i \theta^\prime_{ab}}$ for the off-diagonal elements ($a \ne b$).
%%%%%%%%%%%%%%%%%%%%%
\subsection{Two generation toy model}
In this subsection, we apply the parametrization for $m^{0}_{\nu}$ given in Eq.(\ref{para}) to a two generation toy model of the active neutrinos.
We will carry out the diagonalization of the mass matrix and express the two neutrino masses $m_1$ and $m_2$, the mixing angle $\theta_{12}$, and the Majorana phase $\alpha_{21}$ from Eq.(\ref{eq:2by2PMNS}) in terms of the four invariant parameters in the mass matrix,
\begin{equation}
  m^{0}_{\nu}=\begin{pmatrix} |m_{ee}| & m_{e\mu} \\
    m_{e\mu} & |m_{\mu \mu}| \end{pmatrix},
    \label{mefffortwo}
\end{equation}
where  $m_{e\mu}\equiv |m_{e\mu}|e^{i \theta^\prime_{e\mu}}$.
We assume that the rank of $m^{0}_{\nu}$ is two and the eigenvalues of the Hermite matrix $H^{0}_{\nu}=m^{0 \ast}_{\nu} m^0_{\nu}$ are not degenerate.
If this is the case, the unitary matrix that diagonalizes the Hermite matrix also diagonalizes $m^0_{\nu}$.
The proof of this theorem can be found in Refs.~\cite{Autonne:1915,Takagi:1925}. 
To obtain the unitary transformation, we start from the Hermite matrix $H^{0}_{\nu}$,
\begin{equation}
  H^{0}_{\nu}=m^{0 \ast }_{\nu} m^{0}_{\nu}=\begin{pmatrix}  |m_{ee}|^2+|m_{e \mu}|^2 &   |m_{ee}|m_{e\mu} +m_{e\mu}^\ast |m_{\mu \mu}|   \\  |m_{ee}|m_{e\mu}^\ast +m_{e\mu} |m_{\mu \mu}|    &  |m_{\mu\mu }|^2+|m_{e \mu}|^2  \end{pmatrix}.
%= \begin{pmatrix} h_{11} & h_{12} \\ h^\ast_{12} &h_{22} \end{pmatrix}
\label{H0eff}
\end{equation}
Because the rank of $m^0_{\nu}$ is two, the determinant $\det m^0_{\nu}$ is non-zero.
This is equivalent to the expression $\det H^0_{\nu}>0 $ or,
\begin{equation}
  |m_{ee}|^2 |m_{\mu \mu}|^2 -2|m_{e\mu}|^2 |m_{ee}| |m_{\mu \mu}| \cos2\theta^\prime_{e\mu} +|m_{e\mu}|^4 >0.
\label{eq:rank2}
\end{equation}
Equation \ref{eq:rank2} can be satisfied if either $ \cos2\theta^\prime_{e \mu} \ne 1$ or $|m_{e\mu}|\ne \sqrt{|m_{ee}| |m_{\mu \mu}|}$ holds true.

Next we study the degeneracy of the eigenvalues of $H^{0}_{\nu}$ by obtaining the eigenvalues $\lambda=m^2_1 , m^2_2$ from $\det (H^{0}_{\nu}-\lambda)=0$.  The solutions are,
\begin{gather}
  {m_1}^2=\frac{|m_{ee}|^2+|m_{\mu \mu}|^2+2 |m_{e \mu}|^2- \sqrt{(|m_{ee}|^2-|m_{\mu \mu}|^2 )^2+4|m_{e\mu}|^2u}}{2}, 
  \label{eq:mass1}
\\
  {m_2}^2=\frac{|m_{ee}|^2+|m_{\mu \mu}|^2+2 |m_{e \mu}|^2+\sqrt{(|m_{ee}|^2-|m_{\mu \mu}|^2 )^2+4|m_{e\mu}|^2u}}{2},
  \label{eq:mass2}
\\
  u\equiv |m_{ee}|^2+|m_{\mu \mu}|^2+2 |m_{ee}| |m_{\mu \mu}|\cos 2\theta^\prime_{e\mu},
  \label{eq:u}
\end{gather}
where we have assumed the mass hierarchy is $m_1^2 <m_2^2$.
In order to have non-degenerate eigenvalues, the condition $\Delta m^2_{21} >0 $ must be satisfied.
In other words, the condition,
\begin{equation}
  (|m_{ee}|^2-|m_{\mu \mu}|^2 )^2+4|m_{e\mu}|^2(|m_{ee}|^2+|m_{\mu \mu}|^2+2 |m_{ee}| |m_{\mu \mu}|\cos 2\theta^\prime_{e\mu}) >0,
\label{cond2}
\end{equation}
must be met.
Equation \ref{cond2} is satisfied if either $|m_{ee}| \ne |m_{\mu \mu}|$ or $\cos 2\theta^\prime_{e\mu} \ne -1$ is true, and the two mass eigenvalues are not degenerate. 
If both conditions Eq.(\ref{eq:rank2}) and Eq.(\ref{cond2}) are fulfilled by $m^0_{\nu}$ in Eq.(\ref{mefffortwo}), one can also diagonalize it with the unitary matrix $U$ that diagonalizes $H^{0}_{\nu}$.
This leads to the non-degenerate and positive eigenvalues of $m^{0}_\nu$.

We examine the details of the unitary matrix $U$ that diagonalizes $H^{0}_{\nu}$.
We first parametrize the unitary matrix as,
\begin{equation}
  U=\begin{pmatrix} \cos\theta & -\sin \theta e^{i \phi}  \\ \sin\theta e^{-i \phi} & \cos\theta \end{pmatrix}.
  \label{U}
\end{equation}
Following the procedure given in Eqs(\ref{eq:H0})-(\ref{eq:sin2theta}), we can determine the phases $\phi$ and $\theta$ as,
\begin{gather}
  e^{i \phi}=\frac{(|m_{ee}|+|m_{\mu \mu}|) \cos \theta^\prime_{e\mu}}{\sqrt{u}} +i  \frac{(|m_{ee}|-|m_{\mu \mu}|)\sin \theta^\prime_{e\mu} }{\sqrt{u}} \label{phi},
\\
  \cot 2\theta=\frac{|m_{ee}|^2-|m_{\mu\mu}|^2}{2|m_{e\mu}|\sqrt{u}},  \label{eq:ttheta}
\\
  \cos 2\theta=\frac{|m_{\mu\mu}|^2-|m_{ee}|^2}{\sqrt{ (|m_{ee}|^2-|m_{\mu\mu}|^2)^2+4|m_{e\mu}|^2 u}} ,  \label{eq:ctheta}
\\
  \sin 2\theta=- \frac{2|m_{e\mu}|\sqrt{u}}{\sqrt{ (|m_{ee}|^2-|m_{\mu\mu}|^2)^2+4|m_{e\mu}|^2 u}} <0. \label{eq:stheta}
\end{gather}
After some calculations written in Eq.(\ref{UTmU}) and Eqs.(\ref{Z1}-\ref{Z2}), we can show that $m^0_{\nu}$ is diagonalized by the unitary matrix $U$,
\begin{equation}
  U^T m^0_{\nu} U=\begin{pmatrix} Z_1 & 0 \\ 0 & Z_2\end{pmatrix},
  \label{eq:diagonalization}
\end{equation}
with,
\begin{align}
  Z_1=e^{i (\alpha_1-\phi+\theta^\prime_{e\mu})} m_1, && Z_2=e^{i (\alpha_2+\phi+\theta^\prime_{e\mu})} m_2, 
\end{align}
where,
\begin{gather}
  \alpha_1=\arctan \frac{-2|m_{ee}||m_{\mu \mu}|\sin 2\theta^\prime_{e\mu} }{u-\sqrt{(|m_{ee}|^2-|m_{\mu \mu}|^2)^2+4|m_{e\mu}|^2 u}}, \label{eq:alpha1}
\\
  \alpha_2=\arctan \frac{-2|m_{ee}||m_{\mu \mu}|\sin 2\theta^\prime_{e\mu} }{u+\sqrt{(|m_{ee}|^2-|m_{\mu \mu}|^2)^2+4|m_{e\mu}|^2 u}}.
  \label{eq:alpha2}
\end{gather}
Then one can remove the phases of $Z_i$ ($i=1,2$) by multiplying the diagonal  matrix $P$ around Eq.(\ref{eq:diagonalization}),
\begin{equation}
  PU^T m^0_{\nu} U P=\begin{pmatrix} m_1 & 0 \\ 0 & m_2\end{pmatrix},
\end{equation}
where,
\begin{equation}
  P=\begin{pmatrix} e^{-i \frac{(\alpha_1-\phi+\theta^\prime_{e\mu})}{2}} & 0 \\ 0 & e^{-i \frac{(\alpha_2+\phi+\theta^\prime_{e\mu})}{2}}\end{pmatrix}.
\end{equation}
The Majorana phase and the mixing angle of $U_\text{PMNS}$ in Eq.(\ref{eq:2by2PMNS}) can be found from the matrix $UP$ as,
\begin{align}
  UP & = \begin{pmatrix} \cos\theta & -\sin \theta e^{i \phi}  \\ \sin\theta e^{-i \phi} & \cos\theta \end{pmatrix} P
\\
  & =\begin{pmatrix} e^{i\phi_1} & 0 \\ 0 &e^{i \phi_2} \end{pmatrix} \begin{pmatrix} \cos\theta & \sin \theta   \\ -\sin\theta & \cos\theta \end{pmatrix} \begin{pmatrix} 1 & 0 \\ 0 &e^{-i \frac{\alpha_2-\alpha_1}{2}+i \pi} \end{pmatrix},
\end{align}
where the diagonal phases are,
\begin{align}
  \phi_1=-\frac{\alpha_1-\phi+\theta^\prime_{e\mu}}{2}, && \phi_2=-(\phi+\pi)-\frac{\alpha_1-\phi+\theta^\prime_{e\mu}}{2}.
\end{align}
Because the diagonal phases $\phi_1$ and $\phi_2$ can be removed by rephasing the charged leptons fields, the unitary matrix $U_\text{PMNS}$ in Eq.(\ref{eq:2by2PMNS}) becomes,
\begin{equation}
  U_{PMNS}=\begin{pmatrix}
    \cos \theta_{12} & \sin \theta_{12} e^{i \frac{\alpha_{21}}{2}} \\
    -\sin \theta_{12} & \cos \theta_{12} e^{i \frac{\alpha_{21}}{2}} \\
  \end{pmatrix}
  =\begin{pmatrix} \cos\theta & \sin \theta   \\ -\sin\theta & \cos\theta \end{pmatrix}  
  \begin{pmatrix} 1 & 0 \\ 0 &e^{-i \frac{\alpha_2-\alpha_1}{2}+i \pi} \end{pmatrix}.
\end{equation}
Therefore one can find the mixing angle and the Majorana phase as follows, 
\begin{align}
  \theta_{12}&=\theta, \label{eq:mixingangle}
  \\
  \frac{\alpha_{21}}{2}&=\frac{\alpha_1-\alpha_2}{2}+ \pi.
\label{eq:majoranap}
\end{align}
Equations \ref{eq:mixingangle}-\ref{eq:majoranap} together with Eqs.(\ref{eq:alpha1}-\ref{eq:alpha2}), Eqs.(\ref{eq:mass1}-\ref{eq:u}), and Eqs.(\ref{phi}-\ref{eq:stheta}) explicitly show the Majorana phase, two mass eigenvalues and the mixing angle, in terms of the four parameters in $m^0_{\nu}$.

As an example, we consider a special case where the diagonal elements of $m^0_{\nu}$ are degenerate by setting $|m_{ee}|=|m_{\mu \mu}|\equiv a$ and $|m_{e\mu}|\equiv c$.
In this case, the mixing angle and the Majorana phase are found to be,
\begin{gather}
  \theta = -\frac{\pi}{4},
\\
  \frac{\alpha_{21}}{2} = \frac{1}{2} \arctan ( \frac{ \sgn(c_{\theta'_{e \mu}}) \sin \theta'_{e\mu}}{|\cos \theta'_{e\mu}| +\frac{c}{a}}) -\frac{1}{2} \arctan (\frac{\sgn(c_{\theta'_{e \mu}})  \sin \theta'_{e\mu} }{ |\cos \theta'_{e\mu}| -\frac{c}{a}})+\pi.
\end{gather}
The matrices $H^0_{\nu}$, $U$, and the mass eigenvalues $m_i (i=1,2)$ for this case are listed in Table~\ref{tableI}.
\begin{table}[htbp]
\begin{tabular}{|c|c|c|c|c|} \hline
 &  $m^{0}_{\nu}$ & $H^{0}_{\nu}$ &  $U$ & $m_1 , m_2$ \\  \hline
\begin{tabular}{c} $a = b\ne c$ \end{tabular} 
& $\begin{pmatrix} a & c e^{i \thetap} \\ c e^{i \thetap} & a \end{pmatrix}$ 
& $\begin{pmatrix} a^2+c^2 &  2a c c_{\thetap} \\ 2ac c_{\thetap} & a^2+c^2 \end{pmatrix}$& $\begin{pmatrix} \frac{1}{\sqrt{2}} & \frac{\sgn (c_{\thetap}) }{\sqrt{2}} \\ \frac{-\sgn (c_{\theta'_{e \mu}})}{\sqrt{2}}  & \frac{1}{\sqrt{2}} \end{pmatrix} $ & $\sqrt{a^2+c^2 \mp 2ac |c_{\thetap}|} $ 
 \\  \hline
\end{tabular}
\caption{The Hermite matrix, the mass eigenvalues, and the Unitary matrix for a special case of $m^0_{\nu}$ where $|m_{ee}|$ and $|m_{\mu \mu}|$ are degenerate. $c_{\thetap}\equiv \cos \thetap \ne 0$ and $\sgn$ is a sign function.}
\label{tableI}
\end{table}

%%%%%%%%%%%%%%%%%%%%%%%%%%%%%%%%%%%%%%%%%%%%%%%%%%%%%%%%
\section{CP violation for lepton number asymmetries  in the seesaw model and Majorana phase}\label{sec:CPVseesaw} 
In this section, we will discuss the implications of the low energy Majorana mass matrix on a high-energy theory. Previously, a connection has been discussed for a model with discrete symmetries~\cite{Ahn:2010nw}.

To begin, we consider the type I seesaw model for two generations of active neutrinos and two gauge singlet Majorana neutrinos as a high-energy model that generates the effective Majorana mass matrix of the toy model in section~\ref{sec:Meff}.
We will derive the expression for the CP violating phase $\theta^\prime_{e\mu}$ from the effective Majorana mass matrix, which depends on the CP violating phases of the seesaw model.
This analysis will demonstrate how the single Majorana phase $\alpha_{21}$ is related to the CP violating phases in the seesaw model via Eq.(\ref{eq:majoranap}) and Eqs.(\ref{eq:alpha1}-\ref{eq:alpha2}).
Simultaneously, we will study how the lepton family asymmetries are generated through the decays of the two heavy Majorana neutrinos.
Combined with the analysis above, we will establish how the CP violation of the low energy $\alpha_{21}$ is related to the CP violation for those lepton family asymmetries.

The Lagrangian for lepton sector of the type I seesaw model of for two generations is,
\begin{equation}
  {\cal L}=-\frac{1}{2} \sum_{i=1,2} \overline{(N_{i R})^c} M_{i} N_{i R} -\sum_{i=1,2} \sum_{a=e,\mu} y_{\nu a i} \overline{L_a} \tilde{\phi} N_{i R} - \sum_{a=e,\mu} y_a \overline{L_a} \phi l_{a R}-h.c.,
\end{equation}
where $N_{i R}$ $(i=1,2)$ are two gauge singlet right-handed neutrinos and $L_a=\begin{pmatrix} \nu_{a L} & l_{a L} \end{pmatrix}^T$ $(a=e, \mu)$ denotes the lepton doublet for the electron and muon families.
After integrating out the gauge singlet heavy neutrinos, one obtains a Majorana mass matrix for the left-handed neutrinos in the broken phase of the electroweak symmetry,
\begin{equation}
  \mathcal{L}_{\nu}=-\frac{1}{2} \overline{(\nu_{a L})^c} ({m _{\nu}})_{ab}\nu_{b L}, 
\end{equation}
where,
\begin{equation}
  {(m_{\nu})}_{a b}= -\sum_{i=1,2}  m_{D a i} \frac{1}{M_i} m_{D b i} \label{meffinthe seesaw}. 
\end{equation}
Next we parametrize the Dirac mass matrix $m_{Di}$ $(i=1,2)$ with the normalized Yukawa vector $u_{a i}$ and a normalization constant, 
\begin{equation}
  m_{D a i}=y_{\nu a i} \frac{v}{\sqrt{2}}=u_{a i} m_{Di},
\label{mD}
\end{equation}
such that,  
\begin{align}
  \sum_{a=e,\mu} u^\ast_{a i} u_{a i}=1, && m_{D i}=\sqrt{\sum_{a=e,\mu} m^\ast_{D a i}m_{D a i}}.
\end{align}
Then the effective Majorana mass matrix can be written in the following form~\cite{Fujihara:2005pv},
\begin{equation}
  (m_{\nu})_{a b} = -\sum_{i=1,2}  u_{a i} X_i  u_{ b i},
\end{equation}
with,
\begin{equation}
  X_i=\frac{m^2_{D i}}{M_i}.
\end{equation}
After a suitable rephasing transformation, like Eq.(\ref{weaktr}), the effective Majorana mass matrix becomes $(m^0_{\nu})$ from Eq.(\ref{para}).

For the two generation case, the rephasing transformation leads to a form like Eq.(\ref{mefffortwo}) written as:
\begin{align}
  (m^0_{\nu})_{a b}&= -\sum_{i=1,2} e^{i (\delta_a+\delta_b)} u_{a i} X_i  u_{ b i},
  \\
  &= -\sum_{i=1,2} W_{a i} W_{b i},
  \label{eq:meffW}
\end{align}
where $\delta_a (a=e, \mu)$ are chosen so that the diagonal elements of $(m^0_{\nu})_{a b}$ are real and positive.
The following complex vectors $W_{a i}$ were introducted in Eq.(\ref{eq:meffW}),
\begin{equation}
  W_{a i}=e^ {i\delta_a} u_{a i} \sqrt{X_i}=|W_{a i}| e^{i \frac{\alpha_{a i}}{2}}, \label{W}
\end{equation}
where,
\begin{align}
  \arg W_{a i}\equiv \frac{\alpha_{a i}}{2}, && |W_{a i}|=\sqrt{X_i} |u_{a i}|. 
\end{align}
Using those complex vectors, the diagonal elements of $m^0_{\nu}$ satisfy,
\begin{gather}
  |m_{aa}|+ W_{a1}^2+W_{a2}^2=0, \label{maa}
\\
  |m_{ee}|+ (W_{e1})^2+(W_{e2})^2=0, \label{mee}
\\
  |m_{\mu\mu}|+ (W_{\mu1})^2+(W_{\mu2})^2=0,\label{mmumu}
\end{gather}
while the off-diagonal elements of $m^0_{\nu}$ is are,
\begin{gather}
  m_{e\mu}=-\sum_{i=1}^{2} W_{e i} W_{\mu i},
  \\
  |m_{e\mu}|e^{i \theta'_{e\mu}} = \sum_{i=1}^2 |W_{e i}| |W_{\mu i}| e^{i (\frac{\alpha_{e i}+ \alpha_{\mu i}}{2}-\pi )}.\label{memu}
\end{gather}

\begin{figure}[htb]
\begin{minipage}[b]{0.48\columnwidth}
    \centering
    \includegraphics[width=1.0\columnwidth]{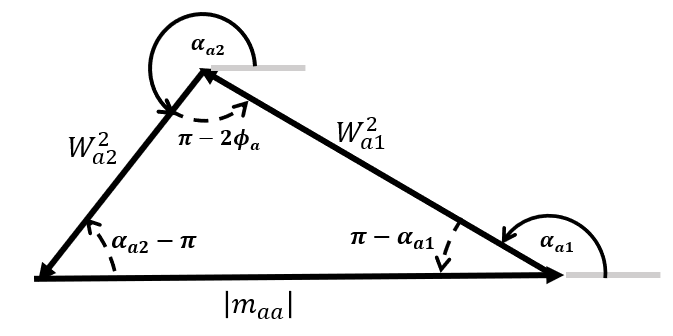}
    \caption{Triangle for the diagonal element of the effective Majorana mass matrix
$|m_{aa}|$ for the type I seesaw model with two right-handed  neutrinos.
The two sides of the triangles represent the contributions from each gauge singlet neutrino. The angles between the two sides  are  related to the CP violating phases $2\phi_a$,$(a=e,\mu)$ for lepton number asymmetries in Eqs.(\ref{asym1}-\ref{asym2}).}
    \label{Figmaatriangle}
\end{minipage}
\hspace{0.04\columnwidth}
\begin{minipage}[b]{0.48\columnwidth}
    \centering
    \includegraphics[width=0.9\columnwidth]{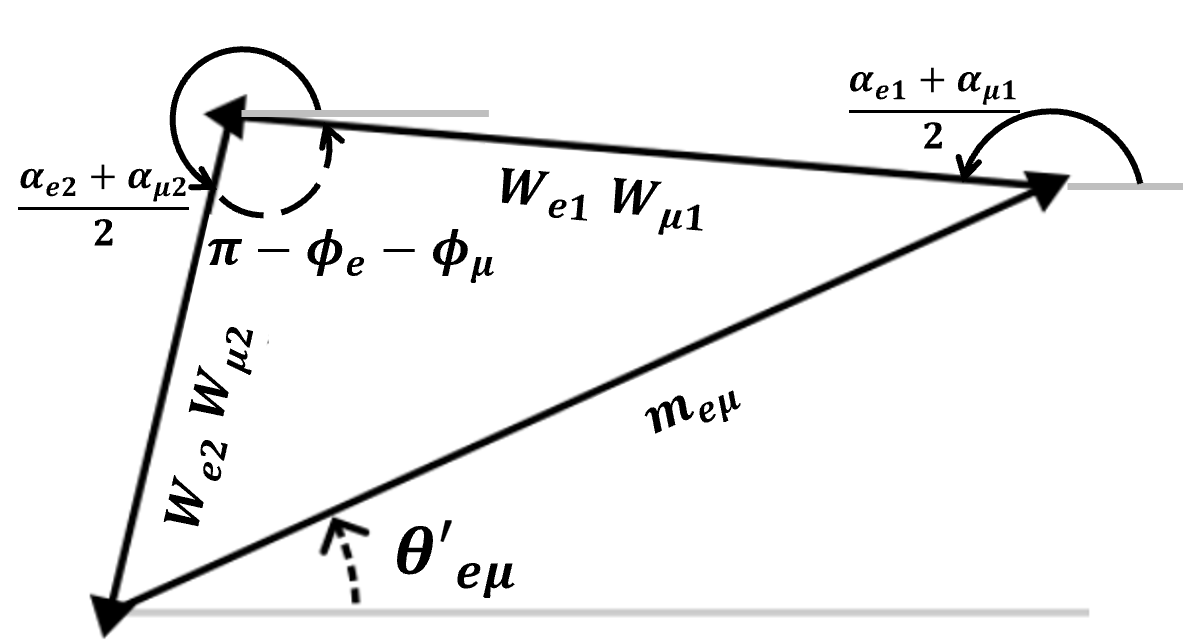}
    \caption{Triangle for the off diagonal element $m_{e\mu}$ of the effective Majorana mass matrix for the type I seesaw model with two right-handed gauge singlet neutrinos. The two sides of the triangle
can be extracted from the analysis of the triangles for the
diagonal elements $|m_{aa}|$ ($a=e, \mu$) in Fig.(\ref{Figmaatriangle}) by obtaining $W_{a i}$ $(a=e,\mu, i=1,2)$.The angle between the two sides in the $m_{e\mu}$ triangle is related to $\phi_e+\phi_\mu$ which also appears
in the lepton number asymmetries in Eqs.(\ref{asym1}-\ref{asym2}).
}
 \label{figmemu}
\end{minipage}
\end{figure}

In Fig.~\ref{Figmaatriangle} and Fig.~\ref{figmemu}, we show the diagonal element of Eq.(\ref{maa}) and off-diagonal element of Eq.(\ref{memu}) in the complex plane.
For an electron family model with two right-handed gauge singlet neutrinos, the triangle for $m_{ee}$ has been drawn in Ref.\cite{YKawakami:2024}.
We can see the CP violating phase at low energy $\theta'_{e\mu}$ can be expressed as,
\begin{equation}
  \begin{split}
    \theta'_{e\mu}&=\arctan \frac{\sum_{i=1}^2 |W_{ei}W_{\mu i}|\sin (\frac{\alpha_{e i}+\alpha_{\mu i}}{2} -\pi)}{\sum_{i=1}^2 |W_{ei}W_{\mu i}|\cos (\frac{\alpha_{e i}+\alpha_{\mu i}}{2} -\pi)},
    \\
    &=\arctan \frac{\sum_{i=1}^2 |u_{ei}u_{\mu i}| X_i \sin (\frac{\alpha_{e i}+\alpha_{\mu i}}{2} -\pi)}{\sum_{i=1}^2 |u_{ei}u_{\mu i}|X_i \cos (\frac{\alpha_{e i}+\alpha_{\mu i}}{2} -\pi)}.
  \end{split}\label{seesawthetap}
\end{equation}
Additionally, the CP violating phase of the seesaw model is characterized by the following relative phases of the two complex numbers $W_{a1}$ and $W_{a2}$.
\begin{gather}
  2 \phi_a=\arg (\frac{W^2_{a2}}{W^2_{a1}})=\arg (\frac{u^2_{a2}}{u^2_{a1}})=\arg(\frac{{m_D}^2_{a2}}{{m_D}^2_{a1}}),
  \label{Aphi}
  \\
  \phi_e+\phi_\mu=\arg(\frac{W_{e2}W_{\mu2}}{W_{e1}W_{\mu1}}),
  \label{phiemu}
\end{gather}
where $a=e,\mu$ is the different families.
We note that these two phases $\phi_a$ are independent of the phases $\delta_a$ in Eq.(\ref{W}), because of the rephasing degrees of freedom.
From Eq.(\ref{Aphi}) the two phases $2\phi_a$ are related to the angle between the two sides of the triangle and $|m_{aa}|$ in Fig.~\ref{Figmaatriangle}.
While Eq.(\ref{phiemu}) tells us that $\phi_e+\phi_\mu$ is related to the angle between the two upper sides of the triangle and $m_{e\mu}$ in Fig.~\ref{figmemu}.

Our goal is to express  $\theta'_{e\mu}$ in terms of $\phi_a$ $(a=e,\mu)$. To help achieve this goal, we note the following relations hold true,
\begin{gather}
  e^{i {(\alpha_{a2}-\pi)}}= \frac{|W_{a2}|^2+|W_{a1}|^2 e^{2i \phi_a}} {|m_{aa}|},
  \\
  e^{i  {(\pi-\alpha_{a1})}}=\frac{|W_{a1}|^2+|W_{a2}|^2 e^{2i \phi_a}} {|m_{aa}|},
  \\
  m_{aa}=\sqrt{|W_{a1}|^4 +|W_{a2}|^4+2 |W_{a1}|^2 |W_{a2}|^2 \cos 2\phi_a}.
\end{gather}
Using those we can derive,
\begin{gather}
  \alpha_{a2}-\pi=\arctan\frac{|W_{a1}|^2 \sin (2 \phi_a)}{|W_{a2}|^2+|W_{a1}|^2 \cos(2 \phi_a)},
  \\
  \pi-\alpha_{a1}=\arctan\frac{|W_{a2}|^2 \sin (2 \phi_a)}{|W_{a1}|^2+|W_{a2}|^2 \cos(2 \phi_a)}.
\end{gather}
Then the combinations of phases $\frac{\alpha_{ei}+\alpha_{\mu i}}{2}$ $(i=1,2)$ that appear in $\theta'_{e\mu}$ of Eq.(\ref{seesawthetap}) are written as,
\begin{align}
  \frac{\alpha_{e2}+\alpha_{\mu 2}}{2}-\pi & = \frac{1}{2} \sum_{a=e}^\mu \arctan\frac{|W_{a1}|^2 \sin (2 \phi_a)}{|W_{a2}|^2+|W_{a1}|^2 \cos(2 \phi_a)},\label{seesawCP1}
\\
  & =\frac{1}{2} \sum_{a=e}^\mu \arctan\frac{|u_{a1}|^2 X_1 \sin (2 \phi_a)}{|u_{a2}|^2 X_2 +|u_{a1}|^2 X_1 \cos(2 \phi_a)},\label{seesawCP1p}
\\
   \pi- \frac{\alpha_{e1}+\alpha_{\mu 1}}{2} & = \frac{1}{2} \sum_{a=e}^\mu \arctan\frac{|W_{a2}|^2 \sin (2 \phi_a)}{|W_{a1}|^2+|W_{a2}|^2 \cos(2 \phi_a)},\label{seesawCP2}
\\
  &=\frac{1}{2} \sum_{a=e}^\mu \arctan\frac{|u_{a2}|^2 X_2 \sin (2 \phi_a)}{|u_{a1}|^2 X_1+|u_{a2}|^2 X_2 \cos(2 \phi_a)}.\label{seesawCP2p}
\end{align}
Equations \ref{seesawCP1}-\ref{seesawCP2p} and Eq.(\ref{seesawthetap}) establish the connection between the CP violating phase at low energy and the CP violation $\phi_a$ $(a=e, \mu)$ of the seesaw model.

It has been proven that the CP violating phase $\phi_a$ also contributes to the lepton family asymmetries of the heavy Majorana neutrinos decays\cite{Fujihara:2005pv,Endoh:2003mz,Covi:1996wh}.
The lepton family aymmetries are given as,
\begin{align}
  \epsilon^{k}_a & = \frac{\Gamma[N^{k}\rightarrow l_a \tilde{\phi}^\dagger ]- \Gamma[N^{k}\rightarrow \overline{l_a} \tilde{\phi} ]}{\sum_{a=e}^{\mu} (\Gamma[N^{k}\rightarrow l_a \tilde{\phi}^\dagger ]+\Gamma[N^{k}\rightarrow \overline{l_a} \tilde{\phi} ])},
\\
  & = \sum_{k' \ne k} \frac{m^2_{D k'}}{4\pi v^2}[\sum_{b=e,\mu}I(x_{k'k}) {\rm Im}(u^\ast_{b k}u_{b k'} u^\ast_{a k}u_{a k'})+\sum_{b\ne a}\frac{1}{1-x_{k'k}}{\rm Im}(u^\ast_{b k}u_{b k'} u_{a k}u^\ast_{a k'})],
\end{align}
where $x_{k'k}=M_{k'}^2 / M_k^2$.
After rewriting the asymmetries with Eq.(\ref{mD}) and Eq.(\ref{Aphi}), one obtains
\begin{gather}
  \epsilon^{1}_a=\frac{m^2_{D_2}}{4\pi v^2}[ I(x_{21}) (|u_{a1}u_{a2}|^2 \sin2\phi_a +|u_{a1}u_{a2} u_{b1} u_{b2}| \sin(\phi_a+\phi_b))+\frac{|u_{a1}u_{a2} u_{b1} u_{b2}| \sin(\phi_a-\phi_b)}{1-x_{21}}], 
  \\
  \epsilon^{2}_a=\frac{-m^2_{D_1}}{4\pi v^2}[ I(x_{12}) (|u_{a1}u_{a2}|^2 \sin2\phi_a +|u_{a1}u_{a2} u_{b1} u_{b2}| \sin(\phi_a+\phi_b))+\frac{|u_{a1}u_{a2} u_{b1} u_{b2}| \sin(\phi_a-\phi_b)}{1-x_{12}}],
\end{gather}
where $a=e, \mu$, $b \ne a$, $|u_{\mu 1}|=\sqrt{1-|u_{e1}|^2}$, and $|u_{\mu 2}|=\sqrt{1-|u_{e2}|^2} $.
The function $I(x)$ is given in Ref.~\cite{Covi:1996wh} as,
\begin{equation}
  I(x)=\sqrt{x} [1+\frac{1}{1-x}+(1+x) \log \frac{x}{1+x}].
\end{equation}
If we sum up the lepton family asymmetries, $\epsilon^{i}=\sum_{a=e}^\mu \epsilon^{i}_a$, one obtains the following expression.
\begin{gather}
  \epsilon^{1}=\frac{m^2_{D_2}}{4\pi v^2}  I(x_{21}) [  |u_{e1}u_{e2}|^2 \sin2\phi_e +2|u_{e1}u_{e2} u_{\mu1} u_{\mu 2}| \sin(\phi_e+\phi_\mu)+|u_{\mu 1}u_{\mu 2}|^2  \sin2\phi_\mu], \label{asym1}
\\
  \epsilon^{2}=\frac{-m^2_{D_1}}{4\pi v^2} I(x_{12}) [ |u_{e1}u_{e2}|^2 \sin2\phi_e +2|u_{e1}u_{e2} u_{\mu1} u_{\mu2}| \sin(\phi_e+\phi_\mu)+|u_{\mu 1}u_{\mu 2}|^2  \sin2\phi_\mu], \label{asym2}
\end{gather}
Then one can find the CP violating phases $\phi_a$ $(a=e, \mu)$ that appear in the lepton number asymmetries \cite{Fukugita:1986hr} of Eqs.(\ref{asym1}-\ref{asym2}) are also related to the Majorana phase at low energies through the relations Eqs.(\ref{eq:majoranap}, \ref{seesawthetap}) and Eqs.(\ref{seesawCP1}-\ref{seesawCP2p}).
%%%%%%%%%%%%%%%%%%%%%%%%%%
\section{Concluding Remarks}
We have investigated an approach to extract the Majorana type-phases of Branco and Rebelo using the time evolution of lepton numbers.
The specific combinations of Majorana type-phases in the same triangle are related to the orientation of unitary triangles for the PMNS matrix, and the Majorana phases $\alpha_{21}$ and $\alpha_{31}$.
After taking the second-order time derivative of the lepton number expectation values, the dependencies on the summation of Majorana type-phases, i.e., $\beta_i+\beta_j$ and $\gamma_i+\gamma_j$, can be determined.
Thus allowing the extraction of the orientation of unitary triangles for the PMNS matrix, and the Majorana phases.

Our result indicates lepton number can theoretically determine the orientation of unitary triangles for the PMNS matrix and the Majorana phases.
Although this could be complimentary to neutrinoless double-beta decay, our result is only theoretical.
Thus, we are actively investigating different experimental possibilities, which we plan to discuss in future work.
In addition to the Majorana phases, we have also shown how the time derivative of the total lepton number is sensitive to the lightest neutrino mass.
The above features are numerically demonstrated for a two generation toy model.
Investigating beyond the toy model would be better suited after experimental possibilities are determined.
For the future, we are interested in identifying a possible experimental setup used to measure the quantities discussed.

Furthermore, within a toy model, we have also investigated the implications of such a measurement on the Majorana phases and neutrino masses in a high-energy model.
By explicitly diagonalizing the 2 by 2 effective Majorana mass matrix at low energies, we have connected the mixing angle, the Majorana phase, and the two masses of the light neutrinos to  the parameters of the effective Majorana mass matrix.
This enabled us to further investigate the impact on the high-energy model which generates the effective Majorana mass matrix.
We considered an example, in the type I seesaw model with two gauge singlet heavy Majorana neutrinos.
The relations between the CP violation of the seesaw model and a low energy Majorana phase were calculated by investigating the polygons (triangles) of all the three elements of the effective mass matrix.
Additionally, we established a relation between the lepton family number CP asymmetries of heavy Majorana neutrinos decays and the
low energy Majorana phase within the toy model.

%%%%%%%%%%%%%%%%%%%%%%%%%%%%%%%%
\subsection*{Acknowledgement}
This work is supported by Japan Society for the Promotion of Science (JSPS) KAKENHI Grant Number JP17K05418 (T.M) and JP21K13923(K.Y).
The part of this work has been presented in Planck 2024. We would like to thank G. C. Branco, M.N. Rebelo, A. Tanaka, R. Niimi, H. Takata and Y. Kawakami for useful discussions.
One author, N.J.B, would like to express thanks to the Japanese government Ministry of Education, Culture, Sports, Science, and Technology (MEXT) for the financial support during the writing of this work.

\begin{appendices} 
\renewcommand{\thesection}{\Alph{section}}
\counterwithin*{equation}{section}
\renewcommand\theequation{\thesection\arabic{equation}}
%%%%%%%%%%%%%%%%%%%%%%%%%%%%%
\label{sec:appendix Diagonalization of 2 by 2 mass matrix}
\section{The details of the calculations leading to Eq.(\ref{eq:diagonalization})}
In this appendix, we show the calculation leading to Eq.(\ref{eq:diagonalization}).
One first obtains eigenvectors for the Hermite matrix in Eq.(\ref{H0eff}),
\begin{equation}
  H_{\nu}^0 = \begin{pmatrix} 
    h_{11}& h_{12} \\
    h_{12}^\ast & h_{22}
  \end{pmatrix},
\label{eq:H0}
\end{equation}
with 
\begin{gather}
  h_{11} = | m_{{ee}} |^2 + | m_{e \mu} |^2,
  \\
  h_{22} = | m_{\mu \mu} |^2 + | m_{e \mu} |^2,
  \\
  h_{12} = | m_{{ee}}  | m_{e \mu} + m_{e \mu}^{\ast} | m_{\mu \mu} |
  = | m_{e \mu} | (| m_{{ee}} | e^{i \theta'_{e \mu}} + | m_{\mu \mu} |
  e^{- i \theta'_{e \mu}}),
  \\
  \arg (h_{12}) = \phi = \arctan \left( \frac{| m_{{ee}} | -
  | m_{\mu \mu} |}{| m_{{ee}} | + | m_{\mu \mu} |} \tan {\theta'}_{e \mu}^0
  \right),
  \\
  |h_{12} | =  \sqrt{u}  | m_{e \mu} |,
\end{gather}
where $u$ is defined in Eq.(\ref{eq:u}).
The eigenvectors for $H_{{\nu}}^0$ are column vectors of the unitary matrix in Eq.(\ref{U}).
Therefore, the first column vector satisfies
\begin{equation}
  \begin{pmatrix}
    h_{11}& h_{12} \\
    h_{12}^\ast& h_{22} \end{pmatrix} 
    \begin{pmatrix} \cos\theta \\
  \sin \theta e^{-i \phi} \end{pmatrix}= \lambda_1 \begin{pmatrix} \cos\theta \\
  \sin \theta e^{-i \phi} \end{pmatrix},
\end{equation}
where $\lambda_1$ is a smaller eigenvalue.
We choose the phase $\phi$ to be the same as the argument of $h_{12}$,
\begin{align}
  e^{i\phi} & = \frac{(| m_{{ee}} | + | m_{\mu \mu} |) \cos \theta'_{e \mu}}{ \sqrt{u}} + i \frac{(| m_{{ee}} | - | m_{\mu \mu} |) \sin \theta'_{e \mu}}{ \sqrt{u}}, \label{eq:phi}
\\
 \tan \phi & = \frac{| m_{{ee}} | - | m_{\mu \mu} |}{| m_{{ee}} | + | m_{\mu \mu} |} \tan \theta'_{e \mu}.
\end{align}
Then the mixing angle $\theta$ satisfies the following equation,  
\begin{gather}
  \cos \theta h_{11}+ \sin \theta |h_{12}|= \lambda_1 \cos \theta,
  \\
  \cos \theta |h_{12}| +\sin \theta h_{22} =\lambda_1 \sin \theta,
\end{gather}
which leads to the solution,
\begin{equation}
  \cot 2 \theta = \frac{| m_{{ee}} |^2 - | m_{\mu \mu} |^2}{2 | m_{e\mu} |  \sqrt{u}} 
  \label{eq:cot2theta}.  
\end{equation}
The eigenvalue is,
\begin{equation}
  \begin{split}
    \lambda_1 & =\cos^2 \theta h_{11}+\sin 2\theta  |h_{12}|+\sin^2 \theta h_{22},
    \\
    & = \frac{h_{11}+h_{22}}{2}+\frac{\sin2\theta}{2}(\cot 2\theta(h_{11}-h_{22})+2|h_{12}|),
    \\
    &= m_1^2,
  \end{split}
\end{equation}
where we have chosen the sign of $\sin2\theta$ to be negative so that $\lambda_1$ becomes the smaller eigenvalue $m_1^2$ in Eq.(\ref{eq:mass1}),
\begin{equation}
  \sin 2 \theta = \frac{- 2 | m_{e \mu} |  \sqrt{u } }{\sqrt{(| m_{{ee}} |^2 - | m_{\mu \mu} |^2)^2 + 4 | m_{e \mu}|^2  u }}  < 0.
  \label{eq:sin2theta}
\end{equation}
Using the unitary matrix $U$ which diagonalizes the Hermite matrix $H^0_{\nu}$, one computes,
\begin{equation}
  U^T m_{{\nu}}^0 U =
  \left(\begin{array}{cc}
   \cos \theta & \sin \theta e^{- i \phi}\\
   - \sin \theta e^{i \phi} & \cos \theta
 \end{array}\right) \left(\begin{array}{cc}
   | m_{{ee}} | & m_{e \mu}\\
   m_{e \mu} & | m_{\mu \mu} |
 \end{array}\right) \left(\begin{array}{cc}
   \cos \theta & - \sin \theta e^{i \phi}\\
   \sin \theta e^{- i \phi} & \cos \theta
 \end{array}\right).
\label{UTmU}
\end{equation}
Then the off-diagonal elements are given by,
\begin{align}
  (U^T m_{{\nu}}^0 U)_{12}& = (U^T m_{{eff}}^0 U)_{21} ,\nn \\
    &= - | m_{{ee}} |  \sin \theta \cos \theta e^{i \phi} + | m_{e\mu} | \cos  2 \theta e^{i \theta'_{e \mu}} + | m_{\mu \mu} | \cos \theta \sin \theta e^{- i \phi},
\end{align}
while the diagonal elements are given as,
\begin{gather}
  (U^T m_{{\nu}}^0 U)_{11}= | m_{{ee}} | \cos^2 \theta + 2 | m_{e \mu} | \sin \theta \cos \theta e^{i (\theta'_{e \mu} - \phi)} + | m_{\mu \mu} | \sin^2 \theta e^{- 2 i \phi},
\\
  (U^T m_{{\nu}}^0 U)_{22}=| m_{{ee}} | \sin^2 \theta e^{2 i \phi} - 2 | m_{e \mu} | \cos \theta \sin \theta e^{i (\phi + \theta'_{e \mu})} + | m_{\mu \mu} | \cos^2\theta.
\end{gather}
On can show the off-diagonal elements vanish,
\begin{align}
  \text{Re} (U^T m_{\nu}^0 U )_{12} &=  | m_{e \mu} | \cos  2
  \theta \cos \theta'_{e \mu} + (| m_{\mu \mu} | - | m_{\text{ee}} |) \cos
  \theta \sin \theta \cos \phi,\nn \\
  & = \sin 2 \theta \left( \cot 2 \theta | m_{e \mu} | \cos \theta'_{e \mu}
  + \frac{1}{2} (| m_{\mu \mu} | - | m_{\text{ee}} |) \cos \phi \right)=0,\\
 \text{Im} (U^T m_{\nu}^0 U )_{12}  &=  | m_{e \mu} | \cos  2 \theta \sin \theta'_{e
  \mu} - (| m_{\mu \mu} |+ m_{\text{ee}} |) \cos \theta \sin \theta \sin \phi, \nn \\
  & = \sin 2 \theta \left( \cot 2 \theta | m_{e \mu} | \sin \theta'_{e
  \mu} - \frac{1}{2} (| m_{\text{ee}} | + | m_{\mu \mu} |) \sin \phi \right)=0,
\end{align}
where we use Eq.(\ref{eq:phi}) and  Eq.(\ref{eq:cot2theta}).
Next we compute the diagonal elements.
\begin{align}
  Z_1&= (U^T m_{\nu}^0 U )_{11}, \nn 
\\
  & = e^{- i \phi} (| m_{\text{ee}} | \cos^2
  \theta e^{i \phi} + 2 | m_{e \mu} | \sin \theta \cos \theta e^{i \theta'_{e
  \mu}} + | m_{\mu \mu} | \sin^2 \theta e^{- i \phi}),\nn
\\
  & = \frac{e^{- i \phi} }{2} \left\{ {| m_{\text{ee}} | e^{i \phi} + | m_{\mu
  \mu} | e^{- i \phi}} +  \sin 2 \theta (2 |  m_{e \mu} |  e^{i \theta'_{e \mu}} + (| m_{\text{ee}} | e^{i \phi} - | m_{\mu \mu} | e^{- i \phi}) \cot 2 \theta) \right \}, \nn
\\
  & = \frac{e^{-i \phi} e^{i \theta'_{e\mu}}}{2\sqrt{u}}\{|m_{ee}|^2+|m_{\mu \mu}|^2-\sqrt{(|m_{ee}|^2-|m_{\mu \mu}|^2)^2 +4 |m_{e \mu}|^2 u} +2 |m_{ee}| |m_{\mu \mu}|e^{-2i \theta'_{e\mu}}\}, \nn
\\
  & = e^{i (\alpha_1-\phi+\theta'_{e\mu})} m_1,
\label{Z1}
\end{align}
where $\alpha_1$ is given in Eq.(\ref{eq:alpha1}) and $m_1^2$ is obtained in Eq.(\ref{eq:mass1}).
The $(22)$ element is also obtained,
\begin{align}
  Z_2 & = (U^T m_{\nu}^0 U )_{22}, \nn
\\
  & = e^{ i \phi} (| m_{\text{ee}} | e^{ i \phi} \sin^2 \theta- 2 | m_{e \mu} | \sin \theta \cos \theta e^{i \theta'_{e\mu}} + | m_{\mu \mu} | \cos^2 \theta e^{- i \phi}), \nn
\\
  & = \frac{e^{ i \phi} }{2} \left\{ {| m_{\text{ee}} | e^{i \phi} + | m_{\mu \mu} | e^{- i \phi}} - \sin 2 \theta (2 |  m_{e \mu} |  e^{i \theta'_{e \mu}} + (| m_{\text{ee}} | e^{i \phi} - | m_{\mu \mu} | e^{- i \phi}) \cot 2 \theta) \right \}, \nn
\\
  & = \frac{e^{i \phi} e^{i \theta'_{e\mu}}}{2\sqrt{u}}\{|m_{ee}|^2+|m_{\mu \mu}|^2+\sqrt{(|m_{ee}|^2-|m_{\mu \mu}|^2)^2 +4 |m_{e \mu}|^2 u} +2 |m_{ee}| |m_{\mu \mu}|e^{-2i \theta'_{e\mu}}\}, \nn
\\
  & = e^{i (\phi+ \theta'_{e\mu}+\alpha_2)} m_2,
\label{Z2}
\end{align}
where $\alpha_2$ is defined in Eq.(\ref{eq:alpha2}) and $m_2$ is given in Eq.(\ref{eq:mass2}).
With $Z_i$ $(i=1,2)$ in Eqs.(\ref{Z1})-(\ref{Z2}), we complete the diagonalization of $m^0_{\nu}$ as shown in Eq.(\ref{eq:diagonalization}).
\end{appendices}

\end{document}